\documentclass[10pt,twocolumn,twoside,dvipsnames]{IEEEtran}
\usepackage{balance}
\usepackage{amsmath, amssymb, amsopn}
\usepackage{amsthm}
\usepackage{graphicx,subfigure}
\usepackage{acronym}
\usepackage{color}
\usepackage{hhline}
\usepackage{booktabs}
\usepackage{array}
\usepackage{multirow}
\usepackage[noadjust]{cite}
\usepackage[dvipsnames]{xcolor}
\usepackage{algorithmic}
\usepackage[linesnumbered,ruled]{algorithm2e}

\let\labelindent\relax

\usepackage{enumitem}

\newcommand{\red}[1]{\textcolor{red}{#1}}
\newcommand{\blue}[1]{\textcolor{blue}{#1}}
\newcommand{\green}[1]{\textcolor{OliveGreen}{#1}}

\newtheorem{theorem}{Theorem}
\newcommand{\argmin}{\operatornamewithlimits{argmin}}
\newcommand{\argmax}{\operatornamewithlimits{argmax}}

\acrodef{AIR}{acoustic impulse response}
\acrodef{ATF}{acoustic transfer function}
\acrodef{CPSD}{cross power spectral density}
\acrodef{DOA}{direction of arrival}
\acrodef{GCC}{generalized cross-correlation}
\acrodef{PCA}{principal source analysis}
\acrodef{PSD}{power spectral density}
\acrodef{RTF}{relative transfer function}
\acrodef{RMSE}{root mean squared error}
\acrodef{RKHS}{reproducing kernel Hilbert space}
\acrodef{SNR}{signal to noise ratio}
\acrodef{SIR}{signal to interference ratio}
\acrodef{SDR}{signal to distortion ratio}
\acrodef{STFT}{short time Fourier transform}
\acrodef{TDOA}{time difference of arrival}
\acrodef{WGN}{white Gaussian noise}
\acrodef{FIR}{finite impulse response}
\acrodef{GSC}{generalized sidelobe canceller}
\acrodef{MAP}{maximum a posteriori probability}
\acrodef{ML}{maximum likelihood}
\acrodef{MUSIC}{multiple signal classification}
\acrodef{ESPRIT}{estimation of signal parameters via rotational invariance}
\acrodef{MMSE}{minimum mean squared error}
\acrodef{GCC-PHAT}{generalized cross-correlation phase transformation}
\acrodef{NN}{nearest-neighbour}
\acrodef{MRL}{Manifold Regularization for Localization}
\acrodef{LCMV}{linearly constrained minimum variance}
\acrodef{mRTF}{mixed-\ac{RTF}}
\acrodef{OMP}{orthogonal matching pursuit}
\acrodef{SPP}{speech presence probability}
\acrodef{EM}{expectation maximization}
\acrodef{BA}{blocking ability}
\acrodef{NPM}{normalized projection measure}
\acrodef{EVD}{eigenvalue decomposition}
\acrodef{NMF}{Non-negative matrix factorization}
\acrodef{BSS}{Blind source separation}
\acrodef{HU}{hyperspectral unmixing}
\acrodef{SPA}{successive projection algorithm}
\acrodef{ICA}{independent component analysis}
\acrodef{MVDR}{minimum variance distortionless response}
\acrodef{TF}{time-frequency}
\acrodef{BASS}{blind audio source separation}

\newcommand\numberthis{\addtocounter{equation}{1}\tag{\theequation}}

\title{Data-Driven Source Separation Based on \\Simplex Analysis}
\author{Bracha~Laufer-Goldshtein,~\IEEEmembership{Student Member,~IEEE}, Ronen~Talmon,~\IEEEmembership{Member,~IEEE}, and Sharon~Gannot,~\IEEEmembership{Senior Member,~IEEE}
\thanks{Bracha~Laufer-Goldshtein and Sharon Gannot are with the Faculty of Engineering, Bar-Ilan University,
	Ramat-Gan, 5290002, Israel (e-mail: Bracha.Laufer@biu.ac.il, Sharon.Gannot@biu.ac.il); Ronen Talmon is with the Viterbi Faculty of Electrical Engineering, The Technion-Israel Institute of Technology, Technion City, Haifa 3200003, Israel, (e-mail: ronen@ee.technion.ac.il).}}

\begin{document}
\maketitle
\sloppy

\begin{abstract}
\ac{BSS} is addressed, using a novel data-driven approach, based on a well-established probabilistic model. The proposed method is specifically designed for separation of multichannel audio mixtures. The algorithm relies on spectral decomposition of the correlation matrix between different time frames. The probabilistic model implies that the column space of the correlation matrix is spanned by the probabilities of the various speakers across time. The number of speakers is recovered by the eigenvalue decay, and the eigenvectors form a simplex of the speakers' probabilities. Time frames dominated by each of the speakers are identified exploiting convex geometry tools on the recovered simplex. The mixing acoustic channels are estimated utilizing the identified sets of frames, and a linear umixing is performed to extract the individual speakers. The derived simplexes are visually demonstrated for mixtures of $2$, $3$ and $4$ speakers. We also conduct a comprehensive experimental study, showing high separation capabilities in various reverberation conditions.
\end{abstract}

\begin{keywords}
\ac{BASS}, \ac{RTF}, spectral decomposition, simplex.
\end{keywords}

\section{Introduction}
\label{sec:inrto}
\acf{BSS} is a core problem in signal processing with numerous applications in various fields, such as: biomedical data processing, audio processing, digital communication, and image processing~\cite{comon2010handbook}. In \ac{BSS} problems, only the output observations are given, whereas neither the original sources nor the mixing systems are known. Separation methods usually rely on some \emph{a priori} hypothesis regarding the characteristics of the original sources or the obtained mixtures.
Assuming that the sources are independent and have non-Gaussian distributions leads to \ac{ICA} methods based on probabilistic or information theoretic criteria~\cite{lee1998independent,hyvarinen2000independent,hyvarinen2004independent}.
\ac{NMF} methods can be employed for signals which admit factorization to non-negative components~\cite{cichocki2009nonnegative}. Sparsity of the signals is also often assumed, allowing a representation as a linear combination of few elementary signals~\cite{zibulevsky2001blind}.

In audio applications, the measured signals in an array of microphones represent convolutive mixtures of the source signals~\cite{makino2007blind,pedersen2008convolutive,vincent2010probabilistic}. The measured signals are obtained by filtering the clean source signals with the corresponding acoustic channels relating the sources and the microphones. The acoustic channels, in a typical reverberant environment, consist of various reflections from the objects and surfaces defining the acoustic enclosure. The measured signals are commonly analysed in the \ac{STFT} domain, where the convolutive mixtures are transformed into multiplicative mixtures at each frequency bin.

\ac{ICA}-based methods can be applied, subject to scale-ambiguity and source permutation problems~\cite{mitianoudis2003audio,sawada2004robust}. Alternatively, numerous separation methods rely on the sparsity of speech sources in the \ac{STFT} domain, assuming that each \ac{TF} bin is occupied by a single source~\cite{yilmaz2004blind}. In algorithms based on \ac{NMF}, the speech spectrum is decomposed to a multiplication of non-negative basis and activation functions~\cite{fevotte2009nonnegative,ozerov2010multichannel}. Due to joint estimation of source parameters and mixing coefficients, these methods are free from permutation alignment problems.
Other full-band approaches cluster the measurements according to \ac{TDOA} estimates or phase difference levels with respect to several microphones~\cite{arberet2010robust,mandel2010model,traa2014multichannel}.
However, these models cannot be successfully applied in the presence of high reverberation, when the \ac{TDOA} estimates are of poor quality. Robustness to room reverberations can be attained by performing bin-wise clustering, in the cost of adding a second stage of permutation alignment procedure~\cite{winter2007map,sawada2011underdetermined}. The TIFROM algorithm~\cite{abrard2005time} avoids the \ac{TF} sparsity assumption. It inspects the variations of computed instantaneous ratios, and detects small regions in the \ac{TF} plane with a single active speaker.

In this paper, we present a novel source separation algorithm, which is specifically applicable to speech mixtures. The key point lies in the spectral decomposition of the correlation matrix between different observations. The justification of the method is based on a probabilistic model, in which each observation consists of different portions of the hidden sources. The relative portion of each source is randomly generated according to the sources' probabilities, which vary from one observation to another. Based on this model, we show that the column space of the correlation matrix is spanned by the probabilities of the different sources. Accordingly, the rank of the correlation matrix equals the number of sources, and its eigenvectors form a simplex of the sources' activity probabilities. The vertices of the simplex correspond to observations dominated by a single source with high probability, facilitating the estimation of the hidden sources.

The applicability of the presented model for blind separation of speech mixtures relies on two main attributes of multichannel audio mixtures. The first is the sparsity of the speech in the \ac{STFT} domain, implying that different time-frames contain different portions of speech components of the different speakers. The second is the fact that in a multichannel framework each speaker is associated with a unique \emph{spatial signature}, manifested in the associated acoustic channel. Applying the above procedure and exploiting convex geometry tools, we can identify frames dominated by a single speaker, enabling estimation of the corresponding acoustic channels. Given the estimated acoustic channels, the individual speakers are extracted using the pseudo-inverse of the acoustic mixing system.

Our method recovers a simplex of the probability of activity of the different sources. Convex geometry tools are more commonly utilized for \ac{HU} in the emerging field of hyperspectral remote sensing~\cite{bioucas2013hyperspectral,ma2014signal}. In those studies, the goal is to identify materials in a scene, using hyperspectral images with high spectral resolution. The work relies on a linear mixing model, where each pixel is modelled as a linear sum of the radiated energy curves of the materials contained in this pixel. The nature of the problem entails a positivity constraint on the weights of the different materials. In addition, the weights must sum to one due to energy conservation. The latter constraint violates the statistical independence assumption, making the application of many standard \ac{BSS} algorithms inappropriate. Alternatively, the above constraints lay the ground for the application of convex geometry tools for \ac{HU}. There was also an attempt to borrow these principles for quasi-stationary sources such as speech sources~\cite{fu2015blind}. In general, it is clear that speech mixtures are not formed as convex mixtures. In~\cite{fu2015blind}, a certain normalization followed by a pre-processing procedure for cross-correlation mitigation, were proposed in order to enforce bin-wise convexity.

It is important to emphasize that the mixture model presented in this paper is fundamentally different from the one used for \ac{HU}. In our model, we recover a simplex of the probability of activity of the different sources, while in \ac{HU} the simplex is formed in the original (often high-dimensional) domain of the mixing systems. In addition, our method also inherently identifies the number of sources in the mixture, whereas \ac{HU} methods generally assume that the number of sources is known. Moreover, in contrast to~\cite{fu2015blind}, we present a full-band approach based on averaging over a large number of frequency bins, which enhances robustness and avoids permutation problems.

The paper is organized as follows. The probabilistic model and its analysis by convex geometry principles are presented in Section~\ref{sec:stat}. The model is applied to speech mixtures and an algorithm for speaker counting and separation is derived in Section~\ref{sec:alg}. Section~\ref{sec:exp} contains an extensive experimental study demonstrating the performance of the proposed method in comparison to several competing methods. Section~\ref{sec:conc} concludes this paper.

\section{Statistical Mixture Model and Analysis}
\label{sec:stat}
We present a general statistical model describing the generation of a collection of observations as mixtures of a set of hidden sources. The observations consist of different portions of each of the sources, where each source occurs with a certain probability. The separation is based on the computation of the correlation matrix defined over the given observations. Based on the spectral decomposition of the correlation matrix, we can identify the number of hidden sources and derive a simplex representation, which relates each observation with its corresponding probabilities. In Section~\ref{sec:alg}, we discuss the relation between this general model and the problem of blind separation of speech mixtures. We use the analogy between the two to derive an algorithm for estimating the number of active speakers and separating them.

\subsection{Mixture Generation}
\label{sec:mix}
Consider $J$ unknown \emph{hidden sources} $\{\mathbf{h}_j\}_{j=1}^J$. The hidden sources are i.i.d. random vectors consisting of $D$ \emph{coordinates}, i.e. $\mathbf{h}_j \in \mathbb{R}^D$, where the $k$th coordinate of the $j$th source is denoted by $h_j(k)$. The hidden sources follow a multivariate distribution with zero-mean and identity covariance matrix, i.e.:
\begin{equation}
E\left\{\mathbf{h}_j\mathbf{h}_j^T\right\}=\mathbf{I}_D
\label{eq:dist}
\end{equation}
where $\mathbf{I}_D$ is the identity matrix of size $D\times D$. The diagonal covariance matrix implies that the coordinates of the hidden sources are assumed to be uncorrelated. It should be noted that the unit variance assumption is used here for the sake of simplicity, and that the following derivation also holds for non-unit and non-constant variance by applying a proper normalization.

Suppose we are given a set of $L$ observations $\{\mathbf{a}(l)\}_{l=1}^L$, also in $\mathbb{R}^D$, which are formed as a combination of the $J$ hidden sources. Each observation $\mathbf{a}(l)$ is assigned with a set of $J$ probabilities $\{p_j(l)\}_{j=1}^J$ summing to one. The vector $\mathbf{a}(l)$ is constructed by $D$ statistically independent lotteries, which are defined by the associated set of probabilities. In each lottery, the value of the $k$th coordinate of $\mathbf{a}(l)$ is chosen as the value of the $k$th coordinate of the $j$th source $h_j(k)$ with probability $p_j(l)$. Accordingly, the $k$th coordinate of the $l$th observation can be written as:
\begin{equation}
a_l(k)=\sum_{j=1}^JI_j(l,k)h_j(k).
\label{eq:alk}
\end{equation}
where $I_j(l,k)$ is an indicator function, which equals $1$ if the $j$th source is chosen and $0$ otherwise, and satisfies:
\begin{align*}
\sum_{j=1}^JI_j(l,k)&=1 \\ \numberthis \label{eq:indi0}
I_j(l,k)I_i(l,k)&=I_j(l,k)\delta_{ij} \nonumber
\end{align*}
where $\delta_{ij}=1$ for $i=j$ and $\delta_{ij}=0$ otherwise. We further assume that the indicator functions of different coordinates and of different frames are mutually independent.

\begin{figure*}[t!]
\centering
\includegraphics[width=0.8\textwidth,height=0.3\textheight]{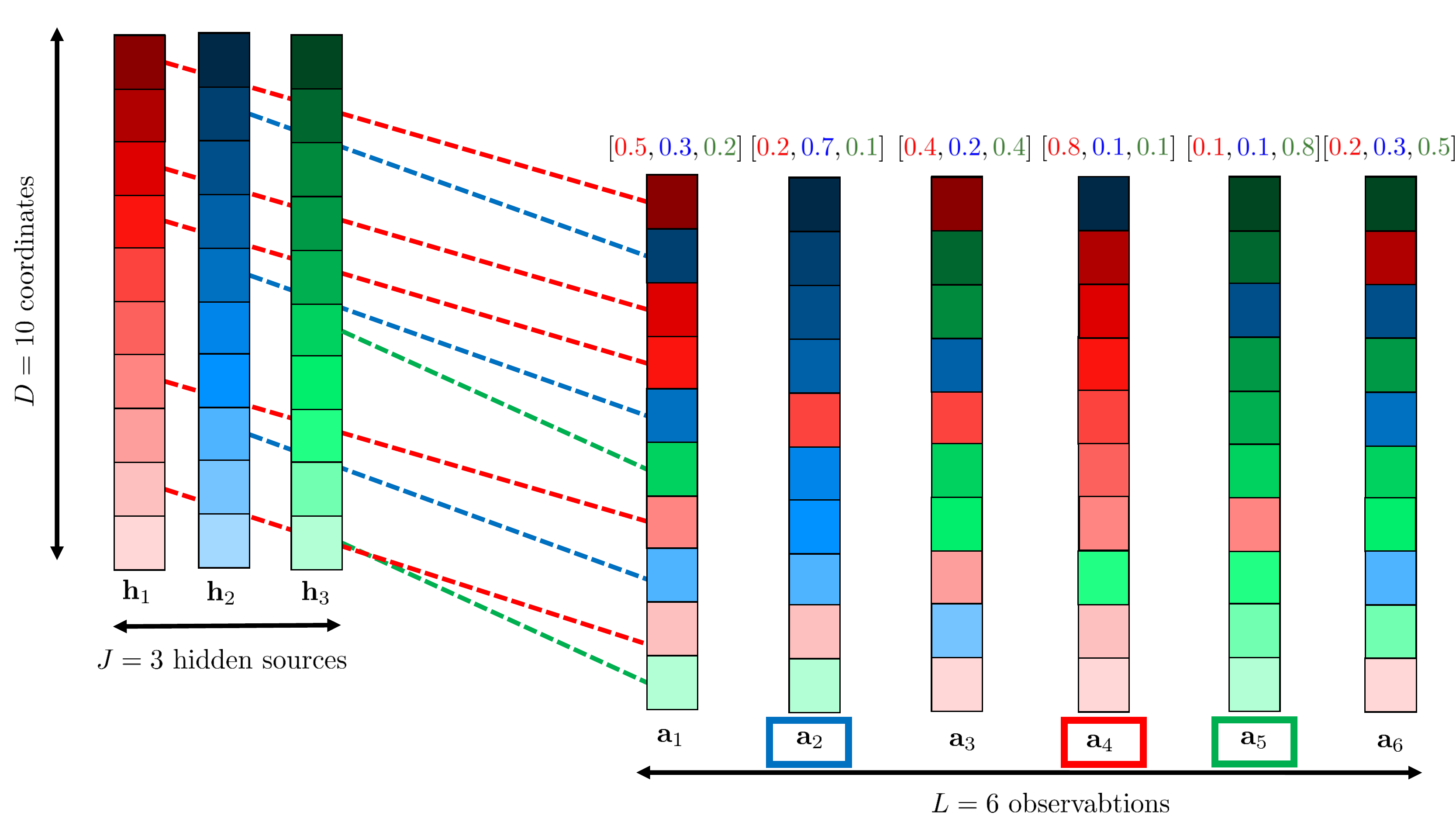}
\centering
\caption{An illustration of the presented statistical mixture model. In this example there are $J=3$ hidden sources $\{\mathbf{h}_j\}_{j=1}^3$ consisting of $D=10$ coordinates, characterized by varying shades of red, blue and green, respectively. The hidden sources are used to construct $L=6$ observations $\{\mathbf{a}(l)\}_{l=1}^6$, where each coordinate is taken from a different source. For the first observation $\mathbf{a}(1)$ dashed lines are drawn between each coordinate and the associated coordinate of the source from which it was taken. The set of probabilities $[\red{p_1(l)},\blue{p_2(l)},\green{p_3(l)}]$ used to construct each observation is written above it. Note that in this example for each observation the number of coordinates taken from each source exactly matches the corresponding probability, while in practice it is only approximately satisfied. Note also that three observations out of the six are highly dominated by a specific source (occupies at least $70\%$ of the observation coordinates). The second observation $\mathbf{a}(2)$ is dominated by the second source $\mathbf{h}_2$. The fourth observation $\mathbf{a}(4)$ is dominated by the first source $\mathbf{h}_1$. The fifth observation $\mathbf{a}(5)$ is dominated by the third source $\mathbf{h}_3$.}
\label{fig:demo}
\end{figure*}

According to this statistical model, for each $l$, the probability $p_j(l)$ corresponds to the \emph{relative portion} of the $j$th source in the construction of the observation $\mathbf{a}(l)$.
An illustration of the presented mixture model is depicted in Fig.~\ref{fig:demo} for $J=3$ sources, $D=10$ coordinates and $L=6$ observations. Consider for example the first observation $\mathbf{a}(1)$, with associated probabilities: $p_1(1)=0.5$, $p_2(1)=0.3$ and $p_3(1)=0.2$. In the vector $\mathbf{a}(1)$, $5$ coordinates are taken from $\mathbf{h}_1$, $3$ coordinates are taken from $\mathbf{h}_2$, and $2$ coordinates are taken from $\mathbf{h}_3$. In practice, the relative portion of each source only approximately matches the corresponding probability for $D$ large enough.

The motivation for this model comes from separation of speech mixtures. According to the sparsity assumption of speech sources in the \ac{STFT} domain~\cite{yilmaz2004blind}, each \ac{TF} bin is dominated by a single speaker. Given the spectrogram of the mixed signal, we can define a column vector for each frame index, consisting of the \ac{STFT} values in a certain frequency band. Relying on the sparsity assumption, each frequency bin in this vector contains a signal from a single speaker. The challenge in speech mixtures, is that they are time-varying. In Section~\ref{sec:mix} we mitigate this problem by proposing features based on the acoustic channels, which are approximately fixed as long as the environment and the source positions do not change dramatically.

\subsection{Analysis of the Correlation Matrix}
\label{sec:corr_anl}
Our goal is to recover the number $J$ of hidden sources $\left\{\mathbf{h}_j\right\}_{j=1}^J$ and to estimate them based on the given set of observations $\{\mathbf{a}(l)\}_{l=1}^L$. The key to our separation scheme lies in the spectral decomposition of the correlation matrix defined over the different observations, which is analysed in this section.

Based on the assumed statistical model~(\ref{eq:dist},~\ref{eq:alk},~\ref{eq:indi0}), the correlation between each two observations $\mathbf{a}(l)$ and $\mathbf{a}(n)$, $1\leq l,n \leq L$ is given by (for details refer to Appendix~\ref{sec:appA}):
\begin{equation}
E\left\{\frac{1}{D}\mathbf{a}^T(l)\mathbf{a}(n)\right\} =
\left\{
	\begin{array}{ll}
		\sum_{j=1}^Jp_j(l)p_j(n) &  \mbox{ if } l\neq n\\
		1 &  \mbox{ if } l = n
	\end{array}
\right..
\label{eq:corrln0}
\end{equation}

\begin{figure*}[t!]
\centering
\subfigure[]{\includegraphics[width=0.32\textwidth,height=0.21\textheight]{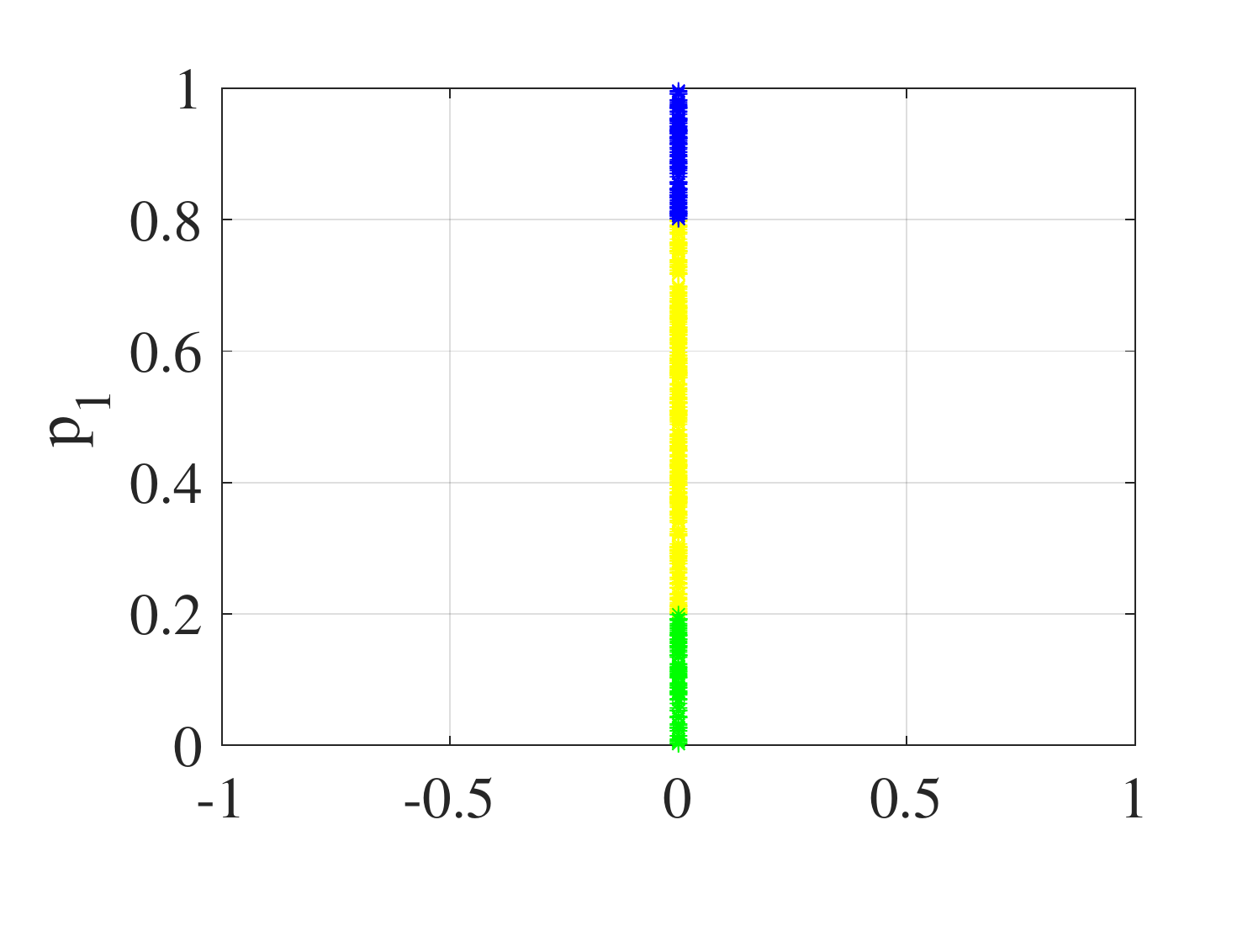}}
\subfigure[]{\includegraphics[width=0.32\textwidth,height=0.21\textheight]{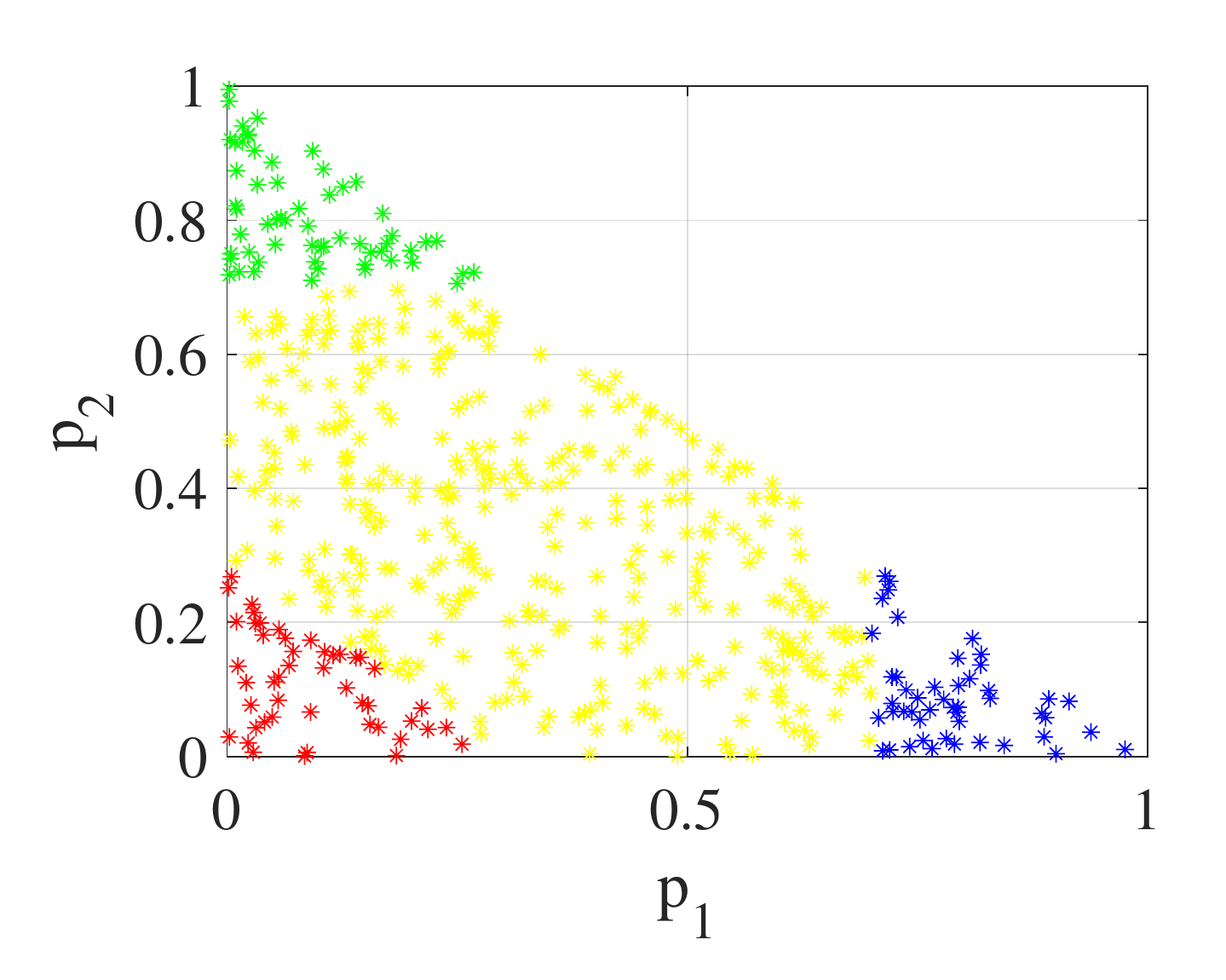}}
\subfigure[]{\includegraphics[width=0.32\textwidth,height=0.21\textheight]{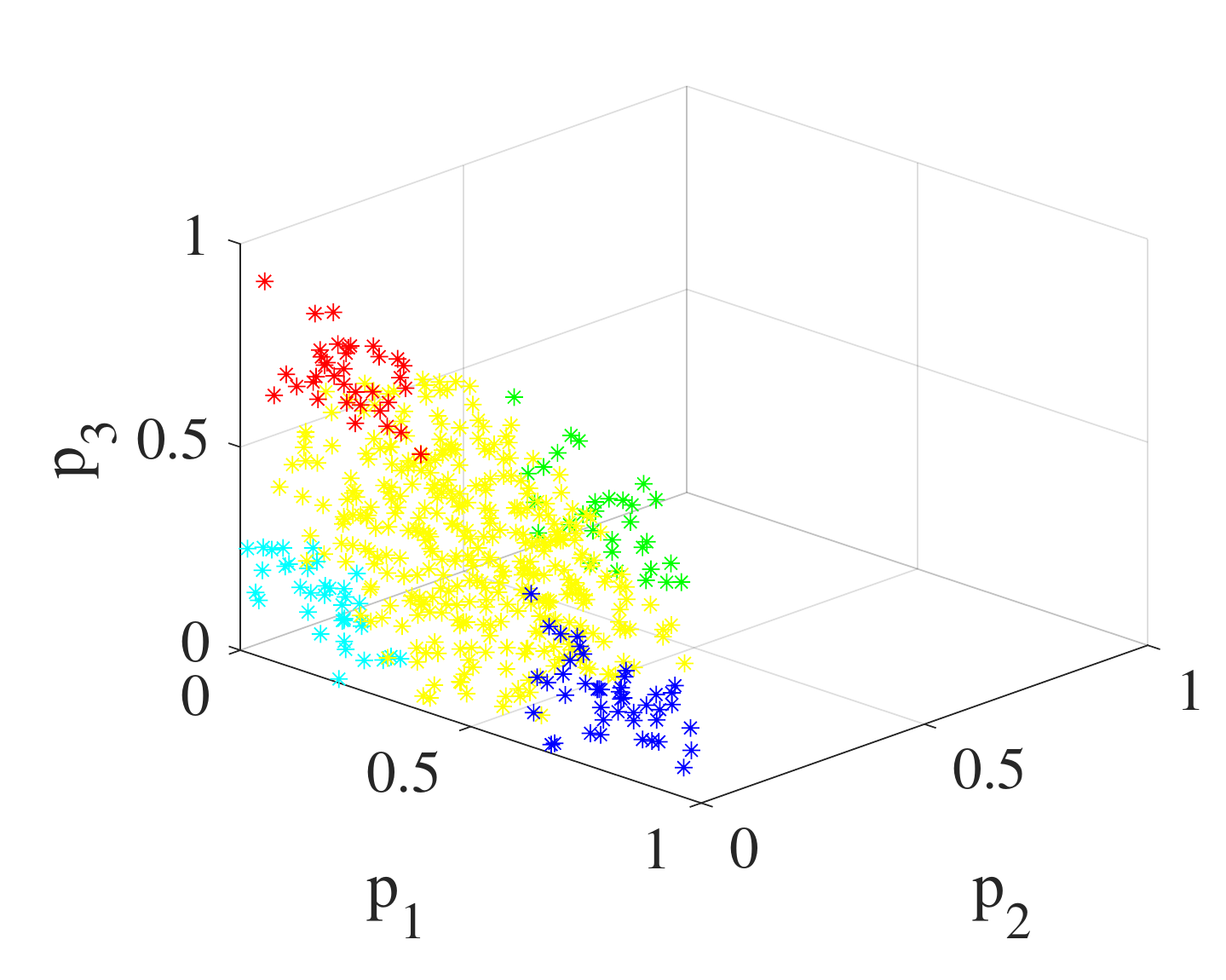}}
\subfigure[]{\includegraphics[width=0.32\textwidth,height=0.21\textheight]{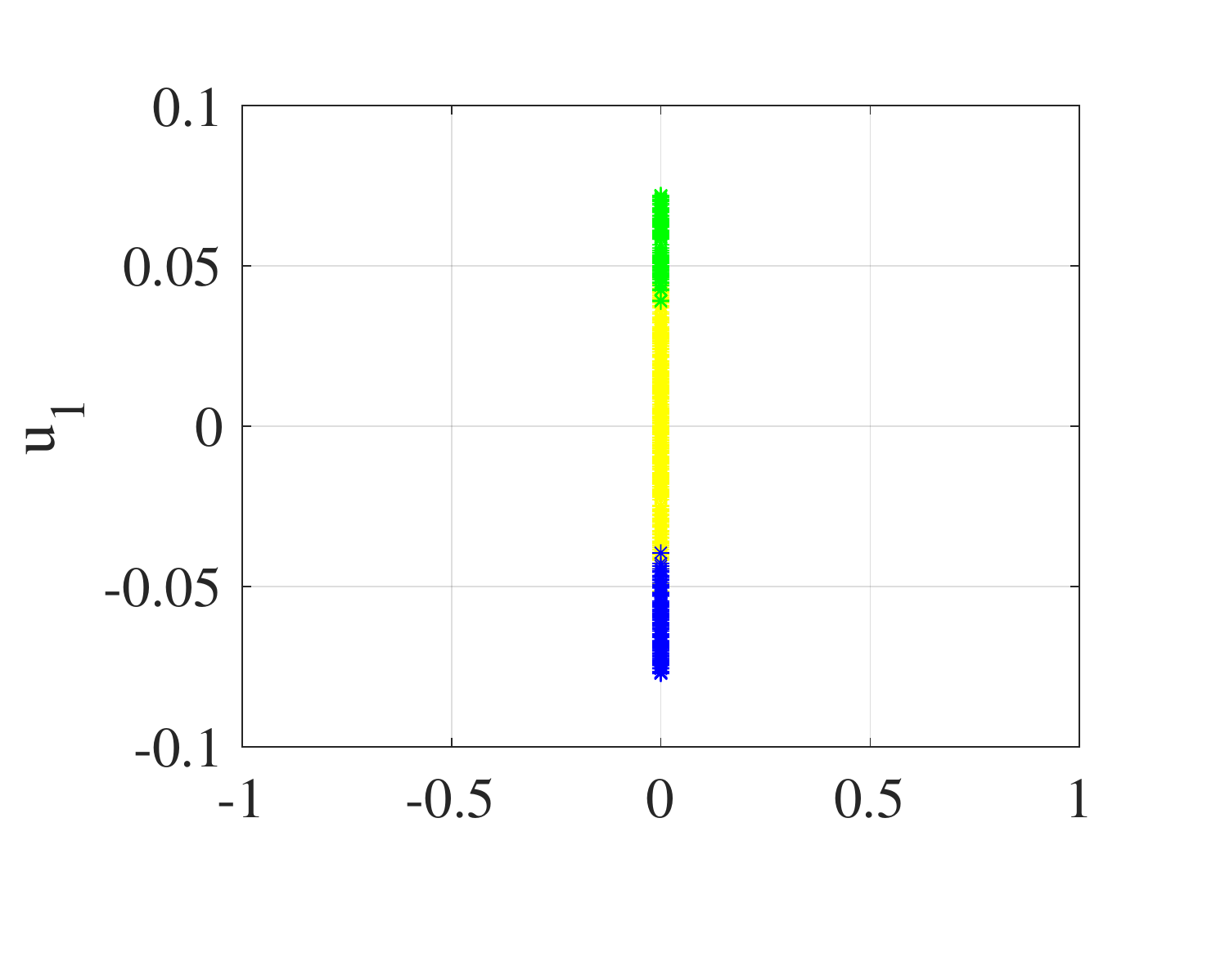}}
\subfigure[]{\includegraphics[width=0.32\textwidth,height=0.21\textheight]{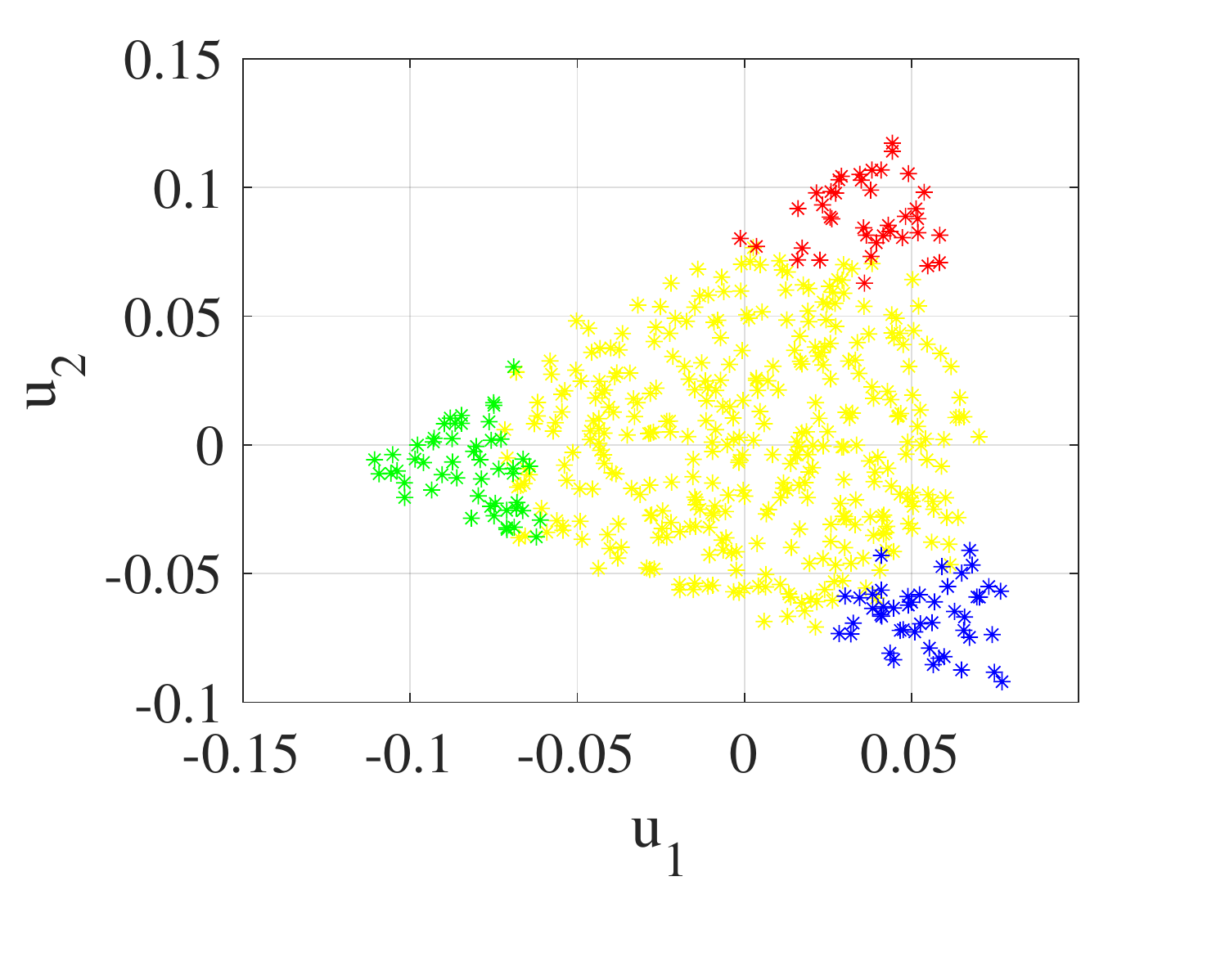}}
\subfigure[]{\includegraphics[width=0.32\textwidth,height=0.21\textheight]{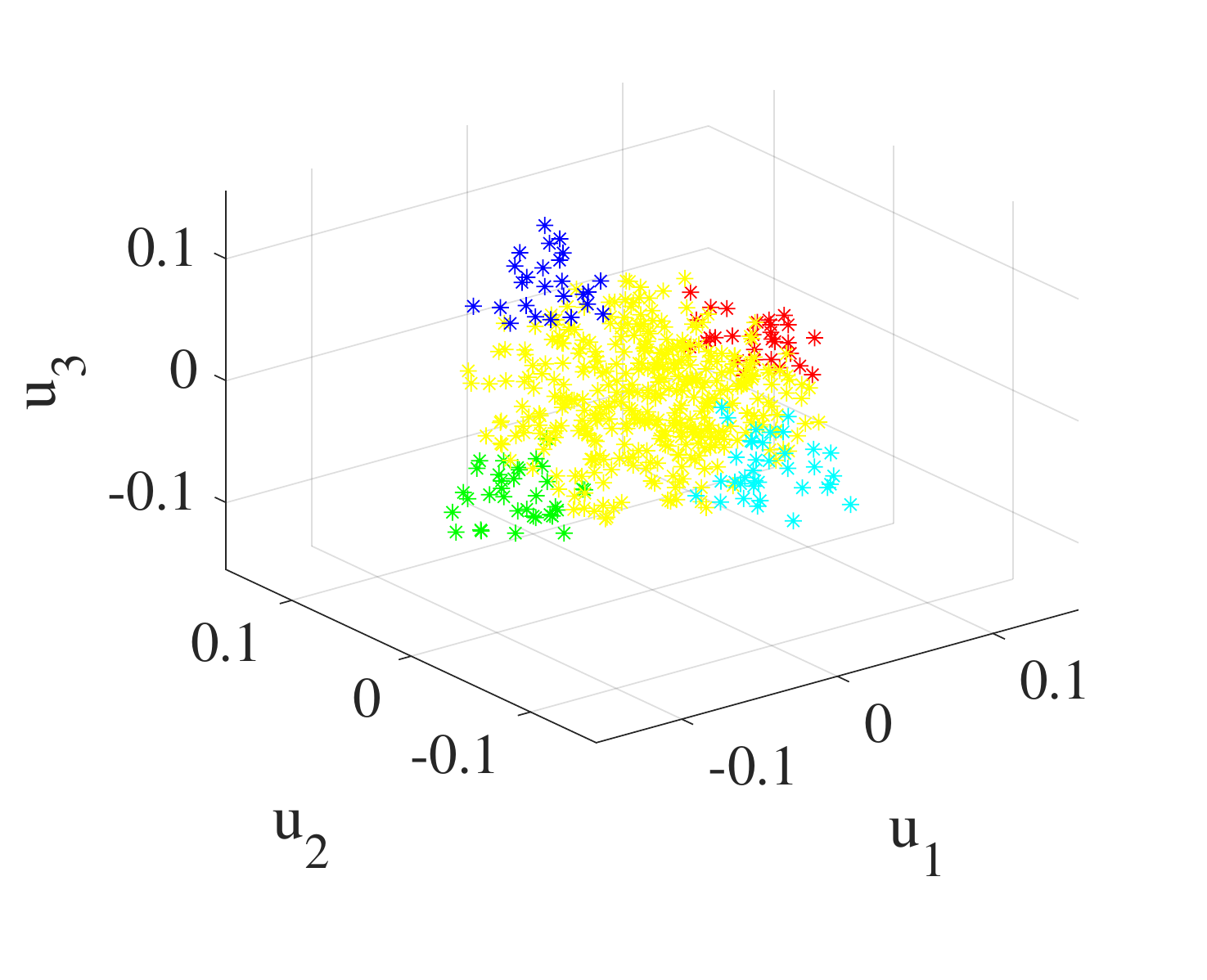}}
\centering
\caption{Scatter plots of $\{\mathbf{p}(l)\}_{l=1}^L$ (oracle probabilities)~(a)-(c) and of $\{\boldsymbol{\nu}(l)\}_{l=1}^L$ (based on the computed eigenvectors)~(d)-(f) obtained for a mixture of $J=2$ (a)~(d), $J=3$ (b)~(e) and $J=4$ (c)~(f) sources. Red, blue, green and cyan points stand for observations dominated by a single source, whereas yellow points stand for observations with multiple sources.}
\label{fig:map_sint}
\end{figure*}

Let $\mathbf{W}$ be the $L\times L$ correlation matrix, with $W_{ln}=E\left\{\frac{1}{D}\mathbf{a}^T(l)\mathbf{a}(n)\right\}$. According to~\eqref{eq:corrln0} the correlation matrix can be recast as:
\begin{equation}
\mathbf{W}=\mathbf{P}\mathbf{P}^T + \Delta \mathbf{W}
\label{eq:PP}
\end{equation}
where $\mathbf{P}$ is a $L\times J$ matrix with $P_{lj}=p_{j}(l)$, and $\Delta \mathbf{W}$ is a diagonal matrix with $\Delta W_{ll}=1-\sum_{j=1}^Jp^2_{j}(l)$. We show in Appendix~\ref{sec:appB}, that $\Delta \mathbf{W}$ has a negligible effect on the spectral decomposition of $\mathbf{W}$. Therefore, henceforth we omit $\Delta \mathbf{W}$ from our derivations and consider the correlation matrix as  $\mathbf{W}\approx\mathbf{P}\mathbf{P}^T$.

Following the mutual independence assumption of the sources, the columns of $\mathbf{P}$ are linearly independent, i.e. the rank of $\mathbf{P}$ equals the number of sources $J$. Hence, the rank of $\mathbf{W}$ also equals $J$, i.e. it has $J$ nonzero eigenvalues. We apply an \ac{EVD} $\mathbf{W}=\mathbf{UD}\mathbf{U}^T$, with $\mathbf{U}$ an orthonormal matrix consisting of the eigenvectors $\{\mathbf{u}_j\}_{j=1}^{L}$, and $\mathbf{D}$ a diagonal matrix with the eigenvalues $\{\lambda_j\}_{j=1}^{L}$ on its diagonal. The eigenvalues $\{\lambda_j\}_{j=1}^{L}$ are sorted by their values in a descending order. According to~\eqref{eq:PP}, the first $J$ eigenvectors $\{\mathbf{u}_j\}_{j=1}^{J}$, associated with the $J$ nonzero eigenvalues $\{\lambda_j\}_{j=1}^{J}$, form a basis for the column space of the matrix $\mathbf{P}$. Accordingly, the following identity holds:
\begin{equation}
\mathbf{U}_\textrm{J}=\mathbf{PQ^T}
\label{eq:proj2}
\end{equation}
where $\mathbf{U}_\textrm{J}=[\mathbf{u}_1,\ldots,\mathbf{u}_{J}]$, and $\mathbf{Q}$ is a $J\times J$ invertible matrix.

Each observation can be represented as a point in $\mathbb{R}^{J}$, defined by the corresponding set of probabilities: $\mathbf{p}(l)=[p_1(l), p_2(l), \ldots, p_J(l)]^T$. Note that each point $\mathbf{p}(l)$ is a convex combination of the standard unit vectors:
\begin{equation}
\mathbf{p}(l)=\sum_{j=1}^{J}p_j(l)\mathbf{e}_j, \> \sum_{j=1}^Jp_j(l)=1.
\end{equation}
where $\mathbf{e}_j=[0,\ldots,1,\ldots,0]^T$ with one in the $j$th coordinate and zeros elsewhere.
Accordingly, the collection of the probability sets $\{\mathbf{p}(l)\}_{l=1}^L$ lies in a $(J-1)$-simplex in $\mathbb{R}^{J}$. This is a \emph{standard simplex}, whose \emph{vertices} are the standard unit vectors $\{\mathbf{e}_j\}_{j=1}^J$.
Note that in this representation, points for which the probability of the $j$th source is dominant over the probabilities of the other sources, i.e. $p_j(l)\gg p_i(l),\> \forall i\neq j, \> 1 \leq i \leq J$, satisfy: $\mathbf{p}(l)\approx \mathbf{e}_j$, namely these points are concentrated nearby the $j$th vertex.

We can use the eigenvectors of $\mathbf{W}$ to form an equivalent representation in $\mathbb{R}^{J}$, defined by: $\boldsymbol{\nu}(l)=[u_1(l), u_2(l), \ldots, u_{J}(l)]^T$. According to~\eqref{eq:proj2}, this representation is related to the former representation by the following transformation:
\begin{equation}
\boldsymbol{\nu}(l)=\mathbf{Q}\mathbf{p}(l).
\label{eq:u}
\end{equation}
Hence, the set $\{\boldsymbol{\nu}(l)\}_{l=1}^L$ occupies a simplex, which is a rotated and scaled version of the standard simplex defined by the standard unit vectors. The new simplex is the convex hull of the following $J$ vertices:
\begin{equation}
\mathbf{e}^*_j=\mathbf{Q}\mathbf{e}_j=\mathbf{Q}_j
\label{eq:Q}
\end{equation}
where $\mathbf{Q}_j$ is the $j$th column of the matrix $\mathbf{Q}$.

Regarding the computation of the matrix $\mathbf{W}$, we do not have access to the expected values $\frac{1}{D}E\{\mathbf{a}^T(l)\mathbf{a}(n)\} , \>\forall 1 \leq l,n \leq L$, hence we use instead the typical values $\widehat{W}_{ln}=\frac{1}{D}\mathbf{a}^T(l)\mathbf{a}(n)$. In Appendix~\ref{sec:appA}, we show that the variance of $\frac{1}{D}E\{\mathbf{a}^T(l)\mathbf{a}(n)\}$ is proportional to $1/D$, hence approaches zero for $D$ large enough, implying that the typical value is close to the expected value.

We demonstrate the above derivation using three examples with $J=2$, $J=3$ and $J=4$ sources. We generate $J$ independent sources of dimension $D=1000$ with $h_j(k)\sim \mathcal{N}(0,1)$
Next, we generate $L=500$ observations, $\{\mathbf{a}(l)\}_{l=1}^L$ according to~\eqref{eq:alk}. To generate the probabilities $\{p_j(l)\}_{j=1}^J$ for each $l$, we draw $J-1$ uniform variables between $[0,1]$ and sort them in an ascending order: $\rho_{1}(l)<\rho_2(l)<\ldots <\rho_{J-1}(l)$. Accordingly, for each $l$, we define the probability of each source by: $p_1(l)=\rho_{1}(l)$, $p_j(l)=\rho_{j}(l)-\rho_{j-1}(l), \> \forall 2\leq j \leq J-1$ and  $p_J(l)=1-\rho_{J-1}(l)$. Next, we construct the matrix $\widehat{\mathbf{W}}$ with $\widehat{W}_{ln}=\frac{1}{D}\mathbf{a}^T(l)\mathbf{a}(n)$, and apply an \ac{EVD}.

Figure~\ref{fig:map_sint}~(a)-(c) depicts $\{\mathbf{p}(l)\}_{l=1}^L$, for $J=2$ (a), $J=3$ (b), and $J=4$ (c). To enable visualization also for $J=4$ we omit one coordinate of $\mathbf{p}(l)$, and represent the simplexes in $\mathbb{R}^{J-1}$. The colouring of the points is as follows: blue, green, red and cyan for observations dominated by the first, the second, the third, and the fourth source, respectively (for $J=3$ only blue, green and red, and for $J=2$ only blue and green).Yellow points depict frames with mixture of sources. We observe that in each plot the points form a $(J-1)$-simplex, i.e. a line segment (a), a triangle (b) and a tetrahedron (c).

Figure~\ref{fig:map_sint}~(d)-(f) depicts $\{\boldsymbol{\nu}(l)\}_{l=1}^L$, for $J=2$ (d), $J=3$ (e), and $J=4$ (f). The coloring of the points is the same as in Fig.~\ref{fig:map_sint}~(a)-(c). We observe that the scattering of the points in~(d)-(f) represents a linear transformation of the scattering in~(a)-(c), as implied by~\eqref{eq:u}.

\begin{figure}[t!]
\includegraphics[width=0.37\textwidth,height=0.18\textheight]{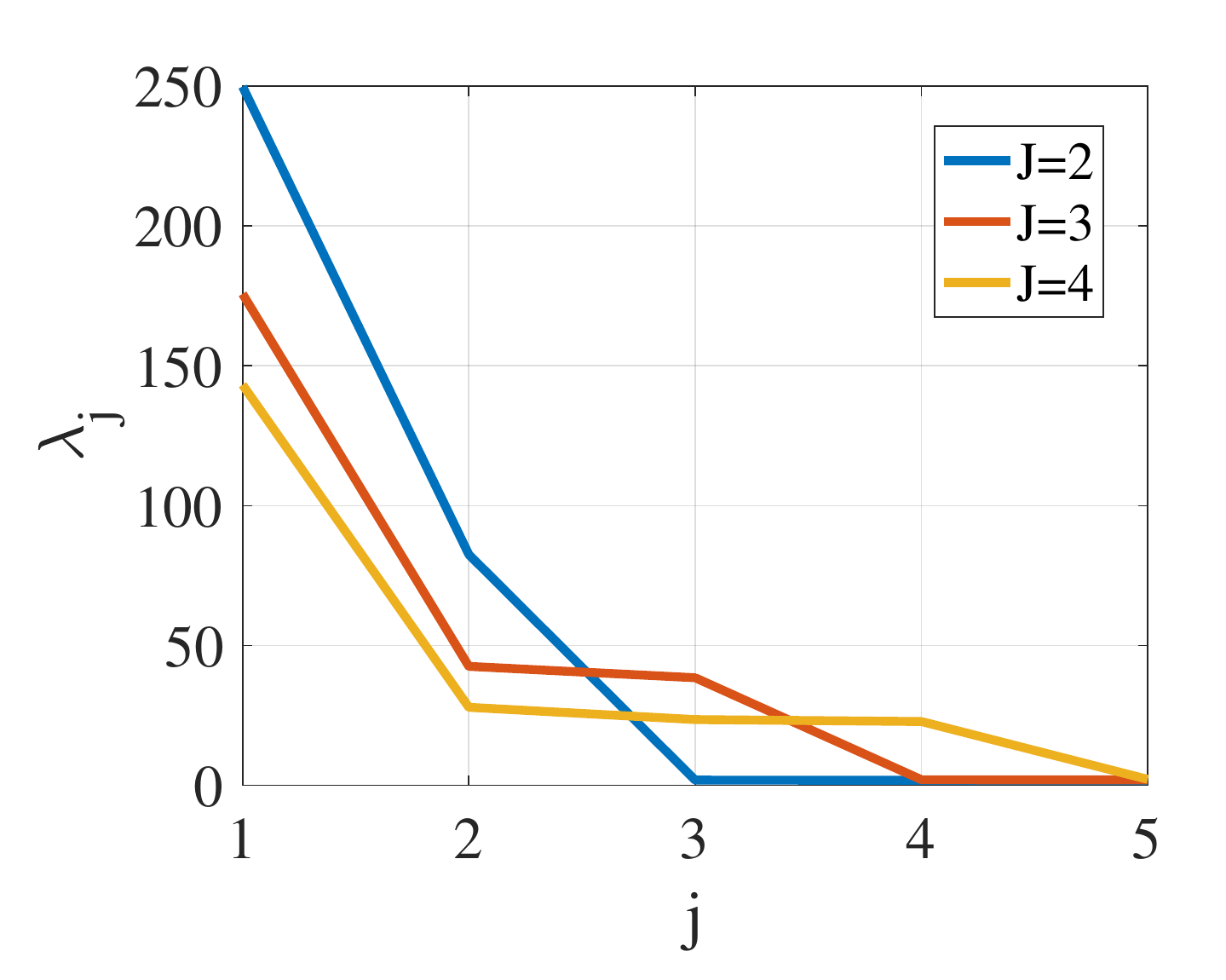}
\centering
\caption{The values of the first $5$ eigenvalues of $\widehat{\mathbf{W}}$, obtained for mixtures with $J=\{2,3,4\}$ sources.}
\label{fig:eig_toy}
\end{figure}

Figure~\ref{fig:eig_toy} depicts the computed eigenvalues of $\widehat{\mathbf{W}}$, sorted in a descending order. We observe that the number eigenvalues with significant value above zero, exactly matches the number of sources $J$.

We conclude with the practical aspects of the new representation derived by the \ac{EVD} of the matrix $\mathbf{W}$. By examining the rank of the obtained decomposition, we can estimate the number of sources involved in the construction of the set $\{\mathbf{a}(l)\}_{l=1}^L$. Furthermore, the eigenvectors $\{\mathbf{u}_j\}_{j=1}^{J-1}$ form a simplex that corresponds to the probability of activity of each source along the observation index $1\leq l \leq L$. We can use this representation to identify observations, which are highly dominated by a certain source, i.e. with $p_j(l)\gg p_i(l), \forall i\neq j$, implying $\mathbf{a}(l)\approx \mathbf{h}_j$. The identified observations can be used for estimating the original $J$ hidden sources $\left\{\mathbf{h}_j\right\}_{j=1}^J$.

\section{Source Counting and Separation}
\label{sec:alg}

In this section, we devise a statistical model for speech mixtures, which resembles the model presented in Section~\ref{sec:mix}. Next, we use the analysis of Section~\ref{sec:corr_anl} to derive an algorithm for source counting and separation.

\subsection{Speech Mixtures}
\label{sec:smix}
Consider $J$ concurrent speakers, located in a reverberant enclosure. The signals are measured by an array of $M$ microphones. The measured signals are analysed in the \ac{STFT} domain with a window of length $N$ samples and overlap of $\eta$ samples:
\begin{equation}
Y^{m}(l,f)=\sum_{j=1}^JY^{m}_j(l,f)=\sum_{j=1}^JA^{m}_j(f)S_j(l,f)
\label{eq:measured}
\end{equation}
where $A^{m}_j(f)$ is the \ac{ATF} relating the $j$th source and the $m$th microphone, and $S_j(l,f)$ is the signal of the $j$th speaker. Here, $f\in\{1,\ldots,K\}$ is the frequency bin, and $l\in\{1,\ldots,L\}$ is the frame index.

The first microphone ($m=1$) is considered as the reference microphone. We define the \acf{RTF}~\cite{gannot2001signal,cohen2004relative} as the ratio between the \ac{ATF} of the $m$th microphone and the \ac{ATF} of the reference microphone, both of which are associated with the $j$th speaker:
\begin{equation}
H^m_j(f)=\frac{A^{m}_j(f)}{A^{1}_j(f)}.
\label{eq:RTF}
\end{equation}
In order to transform the measurements~\eqref{eq:measured} into features that correspond to the model presented in Section~\ref{sec:mix}, we rely on two main assumptions. The first assumption regards the fact that each speaker has a unique spatial signature, which is manifested in the associated \ac{RTF}~\eqref{eq:RTF}. The second assumption regards the sparsity of speech signals in the \ac{STFT} domain.

For speech mixtures, the $J$ hidden sources are defined by the \acp{RTF} of each of the speakers. Each hidden source $\mathbf{h}_j$ consists of $D=2\cdot(M-1)\cdot F$ coordinates for the real and the imaginary parts of the \ac{RTF} values, in $F$ frequency bins and in $M-1$ microphones:
\begin{align*}
\mathbf{h}_j^m&=\left[H^{m}_j(f_1), H^{m}_j(f_2), \ldots, H^{m}_j(f_F)\right]^T\\ \numberthis \label{eq:RTFj}
\mathbf{h}^\textrm{c}_j&=\left[\mathbf{h}_j^{2^T},\mathbf{h}_j^{3^T}, \ldots ,\mathbf{h}_j^{M^T}\right]^T\\
\mathbf{h}_j&=\left[\textrm{real}\left\{\mathbf{h}^\textrm{c}_j\right\}^T,\textrm{image}\left\{\mathbf{h}^\textrm{c}_j\right\}^T\right]^T.
\end{align*}
Note that $\mathbf{h}^1_j$ is an all-ones vector for all $1\leq l \leq L$, hence is excluded from $\mathbf{h}_j$ in~\eqref{eq:RTFj}. We assume that the \ac{RTF} vectors have a diagonal covariance matrix~\eqref{eq:dist}. The attributes  of the Fourier transform prescribe that the real and the imaginary parts of the \ac{RTF} values, as well as the different frequency bins, are uncorrelated. For $F$ large enough, the model can tolerate slight correlations between adjacent frequency bins, or between neighbouring microphones. In addition, we assume that the \acp{RTF} of the different speakers are mutually independent. This was empirically verified  in the experimental study of Section~\ref{sec:exp}, assuming a minimal angle of $30^\circ$ between adjacent speakers.

\begin{figure*}[t!]
\centering
\subfigure[]{\includegraphics[width=0.32\textwidth,height=0.2\textheight]{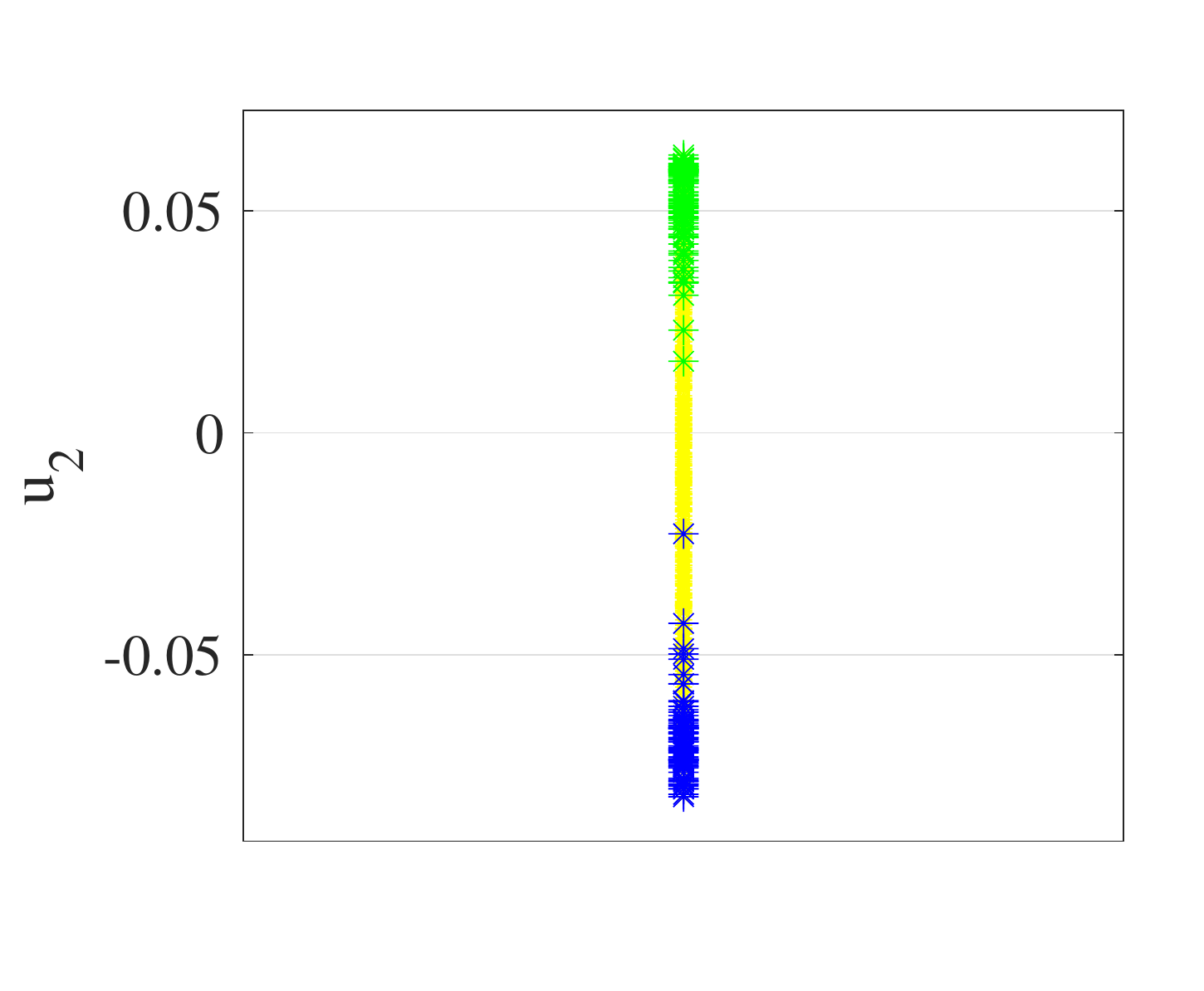}}
\subfigure[]{\includegraphics[width=0.32\textwidth,height=0.19\textheight]{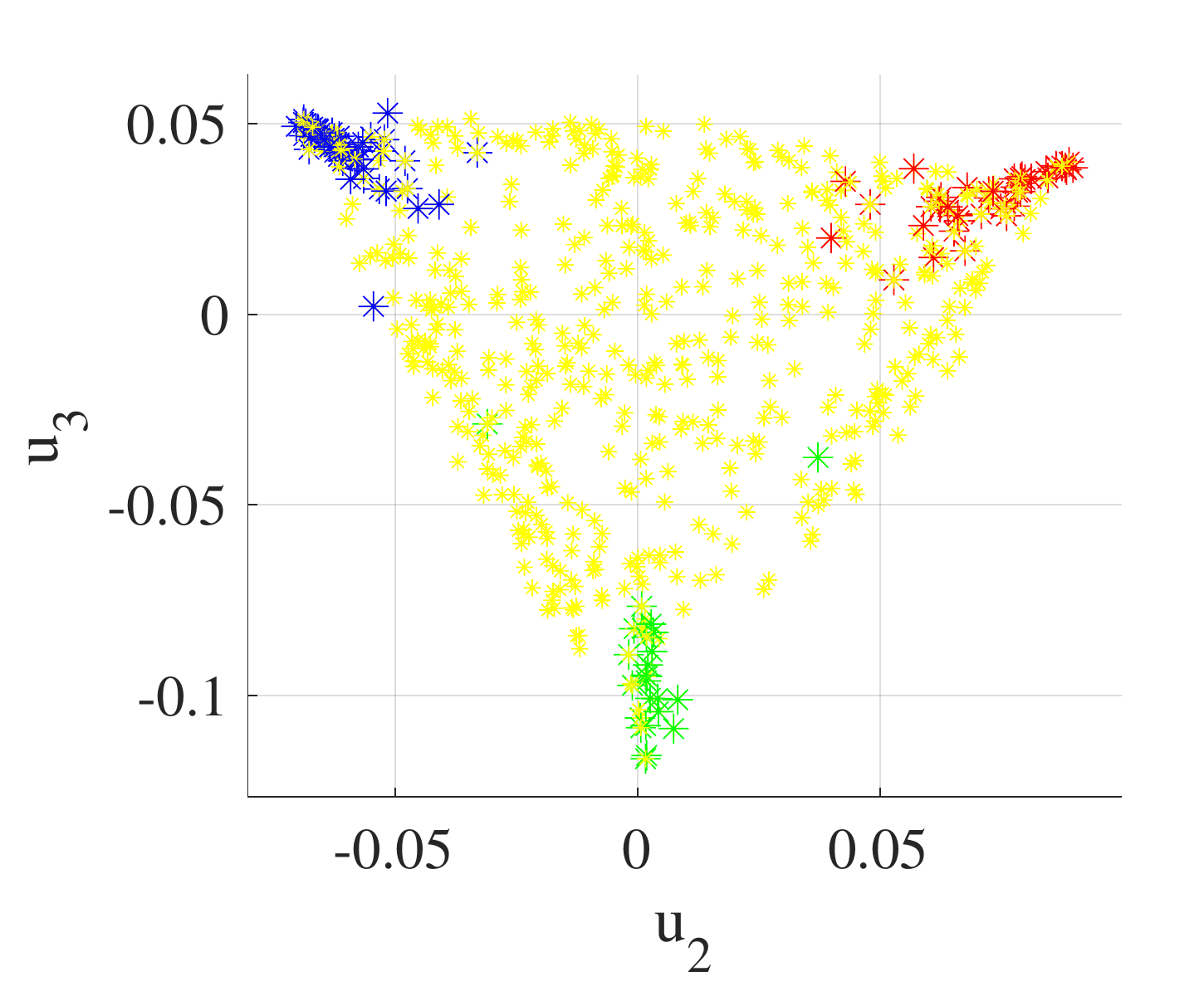}}
\subfigure[]{\includegraphics[width=0.32\textwidth,height=0.2\textheight]{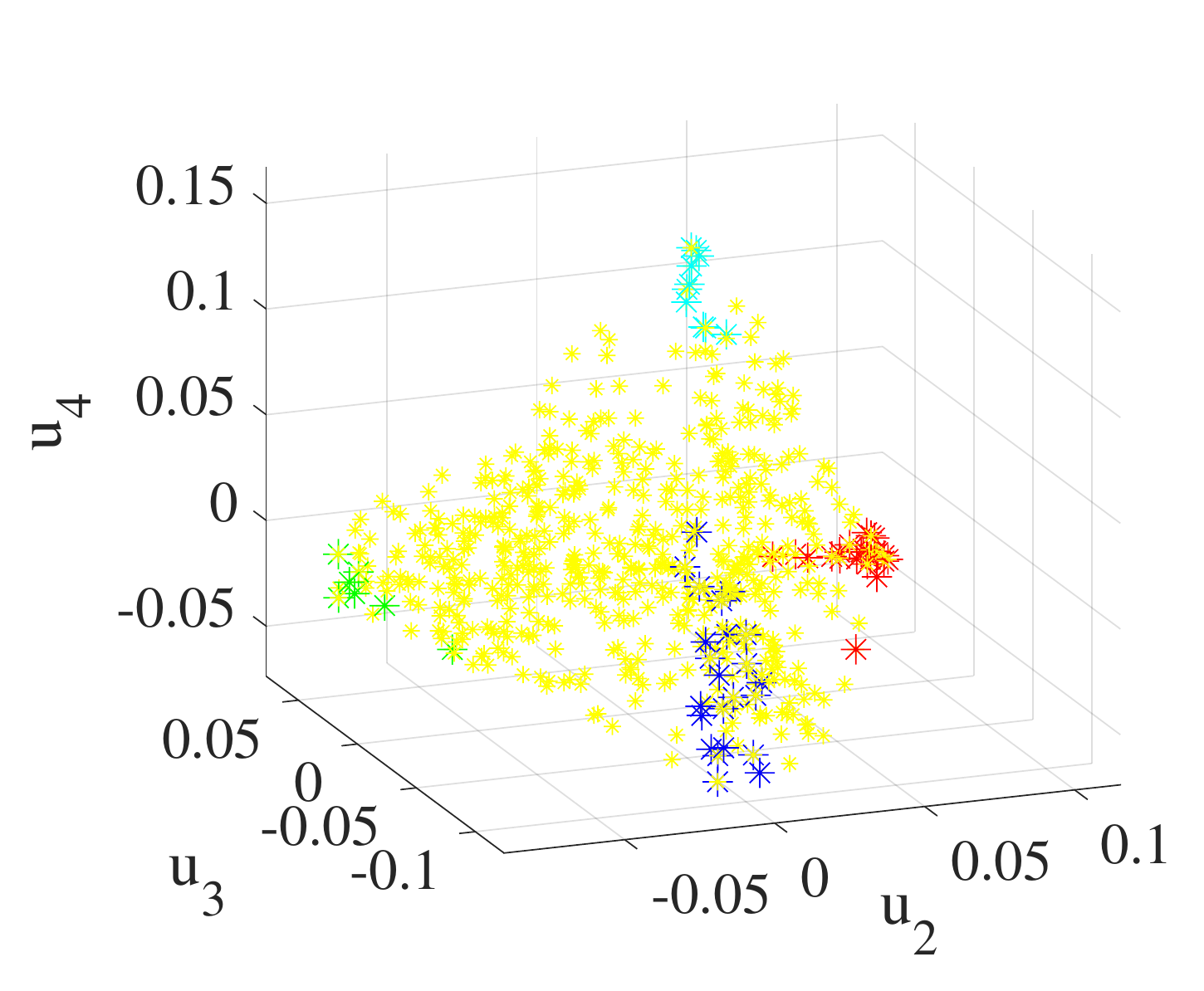}}
\centering
\caption{Scatter plots of $\{\boldsymbol{\nu}(l)\}_{l=1}^L$ for speech mixtures with (a) $J=2$, (b) $J=3$ and (c) $J=4$ speakers.}
\label{fig:map}
\end{figure*}

After defining the $J$ the hidden vectors associated with each of the speakers, we have to extract related observations from the measured signals~\eqref{eq:measured}. We assume that low-energy frames do not contain speech components, and hence these frames are excluded from our analysis. We use the assumption of the speech sparsity in the \ac{TF} domain~\cite{yilmaz2004blind}, which is widely employed in the \ac{STFT} analysis of speech mixtures, and is often applied for localization~\cite{madhu2008scalable,traa2014multichannel,dorfan2015tree} and separation tasks~\cite{ozerov2010multichannel,sawada2011underdetermined,souden2013multichannel}. According to~\cite{yilmaz2004blind}, each \ac{TF} bin is exclusively dominated by a single speaker. Let $I_j(l,f)$ denote an indicator function with expected value $p_j(l)$, which equals $1$ if the $j$th speaker is active in the $(l,f)$th bin, and equals $0$, otherwise. The assumption that the probability $p_j(l)$ is dependent on $l$ but independent of $f$, reflects that the frequency components of a speech signal tend to be activated synchronously~\cite{sawada2011underdetermined,ito2015permutation}. According to the \ac{TF} sparsity assumption, the following holds for each \ac{TF} bin (recall~\eqref{eq:indi0}):
\begin{align*}
\sum_{j=1}^JI_j(l,f)&=1 \\ \numberthis \label{eq:sparse}
I_j(l,f)I_i(l,f)&=I_j(l,f)\delta_{ij} \nonumber
\end{align*}
Hence, ~\eqref{eq:measured} can be recast as:
\begin{equation}
Y^{m}(l,f)=\sum_{j=1}^JI_j(l,f)A^{m}_j(f)S_j(l,f)
\label{eq:indicator}
\end{equation}
We compute the following instantaneous ratio between the $m$th microphone and the reference microphone:
\begin{equation}
R^m(l,k)=\frac{Y^m(l,f)}{Y^1(l,f)}=\frac{\sum_{j=1}^JI_j(l,f)A^{m}_j(f)S_j(l,f)}{\sum_{j=1}^JI_j(l,f)A^{1}_j(f)S_j(l,f)}.
\label{eq:ratio}
\end{equation}
Substituting~\eqref{eq:indicator} and~\eqref{eq:RTF} into~\eqref{eq:ratio}, we get (recall~\eqref{eq:alk}):
\begin{equation}
R^m(l,f)=\sum_{j=1}^JI_j(l,f)H^{m}_j(f)
\label{eq:comb}
\end{equation}
implying that the ratio in the $(l,f)$th \ac{TF} bin equals the \ac{RTF} of one of the speakers. To obtain robustness, we replace the ratio in~\eqref{eq:comb} by power spectra estimates averaged over $T+1$ frames around $l$~\cite{gannot2001signal}:
\begin{equation}
\tilde{R}^m(l,f)\equiv \frac{\hat{\Phi}_{y^{m}y^{1}}(l,f)}{\hat{\Phi}_{y^{1}y^{1}}(l,f)} \equiv \frac{\sum_{n=l-T/2}^{l+T/2}Y^{m}(n,f)Y^{1*}(n,f)}{\sum_{n=l-T/2}^{n+T/2}Y^{1}(n,f)Y^{1*}(n,f)}.
\label{eq:tempRTF}
\end{equation}

Let $\mathbf{a}(l)$ denote the observed \ac{RTF} of frame $l$, which consists of the real and the imaginary parts of the \ac{RTF} values, in $F$ frequency bins and in $M-1$ microphones (recall~\eqref{eq:RTFj}):
\begin{align*}
\mathbf{a}^m(l)&=\left[\tilde{R}^{m}(l,f_1), \tilde{R}^{m}(l,f_2), \ldots, \tilde{R}^{m}(l,f_F)\right]^T\\
\mathbf{a}^\textrm{c}(l)&=\left[\mathbf{a}^{2^T}(l),\mathbf{a}^{3^T}(l), \ldots ,\mathbf{a}^{M^T}(l)\right]^T\\
\mathbf{a}(l)&=\left[\textrm{real}\left\{\mathbf{a}^\textrm{c}(l)\right\}^T,\textrm{image}\left\{\mathbf{a}^\textrm{c}(l)\right\}^T\right]^T. \numberthis
\label{eq:RTFf}
\end{align*}
Note that for a certain frequency bin, the same speaker (both the real and the imaginary parts) is captured by all the microphones. However, this does not affect the relative portions of the different speakers in $\mathbf{a}(l)$, and has a negligible effect on the variance of the correlation~\eqref{eq:var} provided $F\gg M$. There is a trade-off choosing the frequency band $\{f_1,\ldots,f_F\}$. On the one hand, we  should focus on the frequency band in which most of the speech components are concentrated, in order to avoid \ac{TF} bins with low-energy speech components. On the other hand, a sufficient broad frequency band should be used in order to reduce the effect of \ac{TF} bins occupied by several speakers, and to obtain a better averaging with smaller variance~\eqref{eq:var}.

We compute~\eqref{eq:tempRTF} and~\eqref{eq:RTFf} for each $1\leq l \leq L$, and form the set $\left\{\mathbf{a}(l)\right\}_{l=1}^L$. We conclude that the obtained set is constructed from the \ac{RTF} vectors of the different sources~\eqref{eq:RTFj}, and has similar properties to the set of observations defined in Section~\ref{sec:mix}. A nomenclature listing the different symbols and their meanings is given in Table~\ref{tab:symbs}.

\begin{table}[t!]
\centering
\begin{tabular}{ll}
\\
$J$ & No. of sources/speakers, $j\in\{1,\ldots,J\}$ \ \\
$M$ & No. of microphones, $m\in\{1,\ldots,M\}$\\
$L$ & No. of observations/frames in the \ac{STFT}, $l\in\{1,\ldots,L\}$ \\
$F$ & No. of frequency bins in the chosen band, $f\in\{f_1,\ldots,f_F\}$\\
$D$ & No. of coordinates $D=2\times(M-1)\times F$, $k\in\{1,\ldots,D\}$\\
$\mathbf{h}_j$ & Hidden sources defined by \ac{RTF} values of each speaker\\
$\mathbf{a}(l)$ & Observations defined by instantaneous \acp{RTF} of each frame \\
$\mathbf{p}(l)$ & Probability of activity of the speakers in each frame\\
$\mathbf{W}$ & Correlation matrix with $W_{ln}=\frac{1}{D}E\{\mathbf{a}^T(l)\mathbf{a}(n)\}$\\
$\{\lambda_j\}_{j=1}^L$ & Eigenvalues of the correlation matrix $\mathbf{W}$\\
$\{\mathbf{u}_j\}_{j=1}^L$ & Eigenvectors of the correlation matrix $\mathbf{W}$\\
$\boldsymbol{\nu}(l)$& A transformation of $\mathbf{p}(l)$, obtained by the eigenvectors of $\mathbf{W}$\\
$\{\mathbf{e}_j\}_{j=1}^J$ & Vertices of the standard simplex occupied by $\{\mathbf{p}(l)\}_{l=1}^L$\\
$\{\mathbf{e}^*_j\}_{j=1}^J$ & Vertices of the transformed simplex occupied by $\{\boldsymbol{\nu}(l)\}_{l=1}^L$\\
\end{tabular}
\caption{Nomenclature}
\label{tab:symbs}
\end{table}

\subsection{Speaker Counting and Separation}
\label{sec:count_seperate}
After we have shown that the speech separation problem can be formulated using the model in Section~\ref{sec:mix}, we would like to use the analysis of Section~\ref{sec:corr_anl} to derive an algorithm for speaker counting and separation.

Following the derivation of Section~\ref{sec:corr_anl}, we construct an $L\times L$ matrix $\widehat{\mathbf{W}}$ with $\widehat{W}_{ln}= \frac{1}{D}\mathbf{a}^T(l)\mathbf{a}(n)$, and apply \ac{EVD}.
Based on the computed eigenvectors, we form a representation in $\mathbb{R}^J$, defined by: $\boldsymbol{\nu}(l)=[u_1(l), u_2(l), \ldots, u_{J}(l)]^T$.

We provide a similar demonstration for speech mixtures as we have presented in the syntactic case in Section~\ref{sec:corr_anl}. We present three examples with $J=2$, $J=3$ and $J=4$ speakers. The generation of the mixtures and the associated parameters are described in details in the experimental part, in Section~\ref{sec:exp}. Figure~\ref{fig:map} depicts the points $\{\boldsymbol{\nu}(l)\}_{l=1}^L$, for $J=2$ (a), $J=3$ (b) and $J=4$ (c). The plots in Fig.~\ref{fig:map} are generated in a similar way to the plots in Fig.~\ref{fig:map_sint}. Here, too, we omit one coordinate of $\boldsymbol{\nu}(l)$ to enable visualization also for $J=4$.  We observe a good correspondence between Fig.~\ref{fig:map} and Fig.~\ref{fig:map_sint}, which gives evidence to the applicability of the general model of Section~\ref{sec:stat} to the case of speech mixtures.

\begin{figure}[t!]
\centering
\includegraphics[width=0.35\textwidth,height=0.18\textheight]{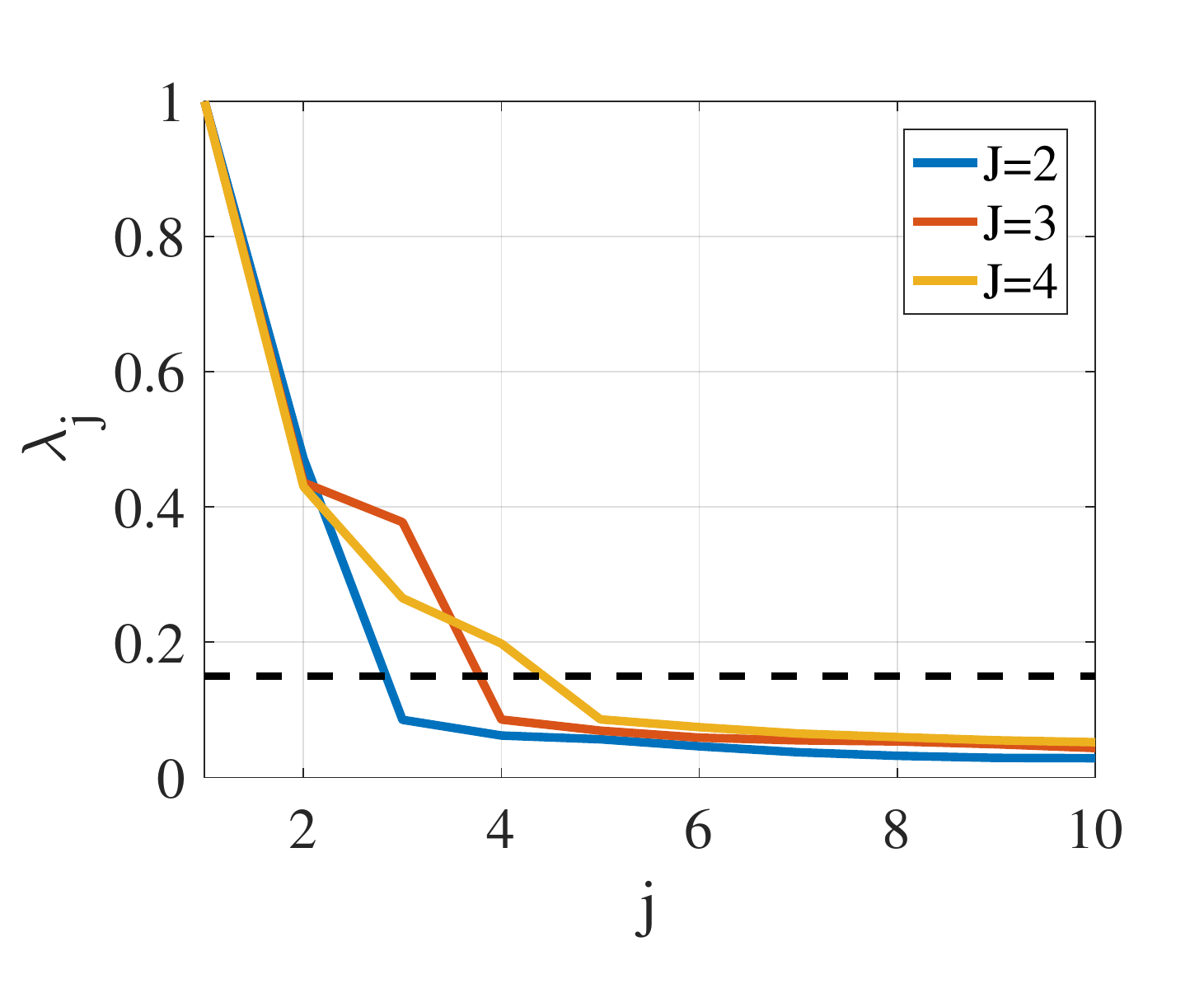}
\centering
\caption{The values of the first $10$ eigenvalues of $\widehat{\mathbf{W}}$, obtained for mixtures with $J=\{2,3,4\}$ speakers.}
\label{fig:eig_real}
\end{figure}

Figure~\ref{fig:eig_real} depicts the computed eigenvalues sorted in a descending order, and normalized by the value of the maximum eigenvalue. As in Fig.~\ref{fig:eig_toy}, the number of eigenvalues with significant value above zero matches the number of sources $J$. Hence, we can estimate the number of sources in the mixture by:
\begin{equation}
\hat{J}=\left(\argmin_{j} \frac{\lambda_j}{\lambda_1}<\alpha\right)-1
\label{eq:Jcov}
\end{equation}
where $\alpha$ is a threshold parameter.

We use the obtained representation $\{\boldsymbol{\nu}(l)\}_{l=1}^L$ to recover the probabilities of the speakers. Next, we detect frames, which are dominated by one of the speakers, and utilize them for estimating the corresponding \acp{RTF}. As discussed in Section~\ref{sec:corr_anl}, the vertices of the simplex defined by $\{\boldsymbol{\nu}(l)\}_{l=1}^L$ correspond to single-speaker points. We recover the simplex vertices, and then utilize them to transform the obtained representation $\{\boldsymbol{\nu}(l)\}_{l=1}^L$ to the original probabilities $\{\mathbf{p}(l)\}_{l=1}^L$.

We assume that for each speaker there is at least one frame, with index $l_j$, which contains only this speaker, i.e. $\mathbf{p}(l_j)=\mathbf{e}_j$. The single-speaker frames are the simplex vertices, i.e. $\boldsymbol{\nu}(l_j)=\mathbf{e}^*_j$. Note that single-speaker frames are tantamount to pure pixels in \ac{HU}. Several algorithms for identifying the vertices of a simplex were developed in the context of \ac{HU}~\cite{boardman1993automating,winter1999n,nascimento2005vertex}. We use a simple approach based on the family of \emph{successive projection} algorithms~\cite{araujo2001successive}. We first identify two vertices of the simplex, and then successively identify the remaining vertices by maximizing the projection onto the orthogonal complement of the space spanned by the previously identified vertices. We start with the first vertex, which is chosen as the point with the maximum norm:
\begin{equation}
\hat{\mathbf{e}}^*_1=\boldsymbol{\nu}(l_1),\> l_1=\argmax_{1\leq l\leq L} \|\boldsymbol{\nu}(l)\|_2
\label{eq:e1}
\end{equation}
Then, the second vertex is chosen as the point with maximum distance with respect to the first identified vertex:
\begin{equation}
\hat{\mathbf{e}}^*_2=\boldsymbol{\nu}(l_2),\> l_2=\argmax_{1\leq l\leq L} \|\boldsymbol{\nu}(l)-\hat{\mathbf{e}}^*_1\|_2.
\label{eq:e2}
\end{equation}

Next, we identify the remaining vertices of the simplex. Let $\bar{\boldsymbol{\nu}}(l)=\boldsymbol{\nu}(l)-\hat{\mathbf{e}}^*_1$ and $\hat{\bar{\mathbf{e}}}^*_j=\hat{\mathbf{e}}^*_j-\hat{\mathbf{e}}^*_1$. Suppose we have already identified $r-1$ vertices $\{\hat{\mathbf{e}}^*_j\}_{j=1}^{r-1}$ with $r>1$. We define the matrix $\mathbf{E}_{r-1}=\left[\hat{\bar{\mathbf{e}}}^*_2,\ldots,\hat{\bar{\mathbf{e}}}^*_{r-1}\right]$, from which we construct its orthogonal complement projector $\mathbf{P}^{\bot}_{r-1}\equiv \mathbf{I}_J-\mathbf{E}_{r-1}(\mathbf{E}_{r-1}^T\mathbf{E}_{r-1})
^{+}\mathbf{E}_{r-1}^T$, where $^+$ denotes the matrix pseudoantique. The $r$th vertex is chosen as the point with maximum projection to the column space of  $\mathbf{P}^{\bot}_{r-1}$:
\begin{equation}
\hat{\mathbf{e}}^*_r=\boldsymbol{\nu}(l_r),\> l_r=\argmax_{1\leq l\leq L} \|\mathbf{P}^{\bot}_{r-1}\bar{\boldsymbol{\nu}}(l)\|_2.
\label{eq:e3}
\end{equation}
We successively repeat~\eqref{eq:e3} for $3\leq r\leq J$, and recover all the simplex vertices $\{\hat{\mathbf{e}}^*_j\}_{j=1}^J$. For simplicity of notation, we ignore possible permutation of the indices of the vertices with respect to the actual identity of the speakers.

Based on~\eqref{eq:Q}, an approximation of the matrix $\mathbf{Q}$ is formed by the identified vertices: $\hat{\mathbf{Q}}=\left[\hat{\mathbf{e}}^*_1,\hat{\mathbf{e}}^*_2,\ldots,\hat{\mathbf{e}}^*_J\right]$. Using the recovered matrix $\hat{\mathbf{Q}}$ we can map the new representation to the original probabilities by (recall~\eqref{eq:u}):
\begin{equation}
\hat{\mathbf{p}}(l)=\hat{\mathbf{Q}}^{-1}\boldsymbol{\nu}(l)
\label{eq:p}
\end{equation}

Let $\mathcal{L}_{j}$ denote the set of frames dominated by the $j$th speaker. Based on the recovered probabilities, we define the set $\mathcal{L}_{j}$ by:
\begin{equation}
\mathcal{L}_{j}=\left\{l\>|\>\hat{p}_j(l)>\beta,\> l\in\{1,\ldots, L\}\right\}
\label{eq:Lj}
\end{equation}
where $\beta$ is a probability threshold.

Given the set $\mathcal{L}_{j}$, an \ac{RTF} estimator of the $j$th speaker, is given by:
\begin{equation}
\hat{H}^m_j(f)=\frac{\sum_{l\in \mathcal{L}_j}Y^m(l,f)Y^{1*}(l,f)}{\sum_{l\in \mathcal{L}_j}Y^1(l,f)Y^{1*}(l,f)}
\label{eq:RTFest}
\end{equation}

Based on the estimated \acp{RTF} $\hat{H}^m_j(k)$ of each of the speakers $1\leq j \leq J$, the mixture can be unmixed applying the pseudo-inverse of the matrix containing the estimated \acp{RTF}:
\begin{equation}
\mathbf{z}(l,f)=\mathbf{B}^H(f)\mathbf{y}(l,f)
\label{eq:unmix}
\end{equation}
where
\begin{align*}
\mathbf{y}(l,f)&=\left[Y^1(l,f),Y^2(l,f),\ldots,Y^M(l,f)\right]^T\\ \numberthis
\mathbf{b}(f)&=\mathbf{C}(f)(\mathbf{C}(f)^H\mathbf{C}(f))^{-1}
\end{align*}
and $\left[\mathbf{C}(f)\right]_{(m,j)}=\hat{H}^m_j(f)$. The time-domain separated signals are obtained by applying the inverse-\ac{STFT}. The proposed method is summarized in Algorithm~\ref{alg:algorithm1}.

\begin{algorithm}
\caption{Separation Algorithm}
\begin{algorithmic}
\STATE \textbf{Feature Extraction:}
\begin{itemize}
\item Estimate instantaneous \acp{RTF} $\left\{\tilde{R}^m(l,f)\right\}_{l,f,m}$~\eqref{eq:tempRTF}.
\item Construct observation vectors $\{\mathbf{a}(l)\}_{l=1}^L$~\eqref{eq:RTFf}.
\end{itemize}
\STATE \textbf{Form a Data-Driven Simplex:}
\begin{itemize}
\item Estimate the correlation matrix $\widehat{\mathbf{W}}$ with $\widehat{W}_{ln}=\frac{1}{D}\mathbf{a}^T(l)\mathbf{a}(n)$.
\item Compute \ac{EVD} of $\widehat{\mathbf{W}}$ and obtain $\{\mathbf{u}_j,\lambda_j\}_{j=1}^L$.
\item Estimate the number of speakers $\hat{J}$~\eqref{eq:Jcov}.
\item Construct $\boldsymbol{\nu}(l)=[\mathbf{u}_1(l),\mathbf{u}_2(l),\ldots,\mathbf{u}_J(l)]$.
\item Form the set $\{\boldsymbol{\nu}(l)\}_{l=1}^L$ lying in a simplex.
\end{itemize}
\STATE \textbf{Recover Activity of Speakers:}
\begin{itemize}
\item Recover simplex vertices $\{\hat{\mathbf{e}}^*_j\}_{j=1}^J$~\eqref{eq:e1},\eqref{eq:e2},\eqref{eq:e3}.
\item Estimate speakers' probabilities $\{\mathbf{p}(l)\}_{l=1}^L$~\eqref{eq:p}.
\item Identify single-speaker frames $\{\mathcal{L}_j\}_{j=1}^J$~\eqref{eq:Lj}.
\end{itemize}
\STATE \textbf{Unmixing Procedure:}
\begin{itemize}
\item Estimate the \acp{RTF} $\left\{\hat{H}^m_j(f)\right\}_{f,j,m}$.
\item Separate the individual speakers~\eqref{eq:unmix}.
\end{itemize}
\end{algorithmic}
\label{alg:algorithm1}
\end{algorithm}

\section{Experimental Study}
\label{sec:exp}
In this section, we evaluate the performance of the proposed method in various test scenarios. The measured signals are generated using concatenated TIMIT sentences. The clean signals are convoluted with \acp{AIR}, which are drawn from an open database~\cite{hadad2014multichannel}. The \acp{AIR} in the database were measured in a reverberant room of size $6$m$\times 6$m$\times 2.4$m with reverberation times of $160$ms, $360$ms and $610$ms. We use a uniform linear array of $M=8$ microphones with $8$cm inter-microphone spacing. The different speaker positions are located on a spatial grid of angles ranging from $-90^\circ$ to $90^\circ$ in $15^\circ$ steps with $1$m and $2$m distance from the microphone array.

The signal duration is $20$s, with sampling rate of $16$kHz. The window length of the \ac{STFT} is set to $N=2048$ with $\eta=75\%$ overlap between adjacent frames, which corresponds to a total amount of $L=622$ frames. For each frame, the instantaneous \ac{RTF} of each frequency bin in~\eqref{eq:tempRTF}, is estimated by averaging the signals in $3$ adjacent frames ($T=2$). The instantaneous \ac{RTF} vectors in~\eqref{eq:RTFf} consist of $F=576$ frequency bins, corresponding to $0-4.5$kHz, in which most of the speech components are concentrated. The obtained concatenated vectors of length $D=2\cdot(M-1)\cdot F=8064$ are normalized to have a unit-norm. The results are demonstrated for mixtures of $J=2$, $J=3$ and $J=4$ speakers in different locations (with a minimum angle of $30^\circ$ between adjacent speakers).

\begin{figure}[t!]
\centering
\subfigure[]{\includegraphics[width=0.35\textwidth,height=0.18\textheight]{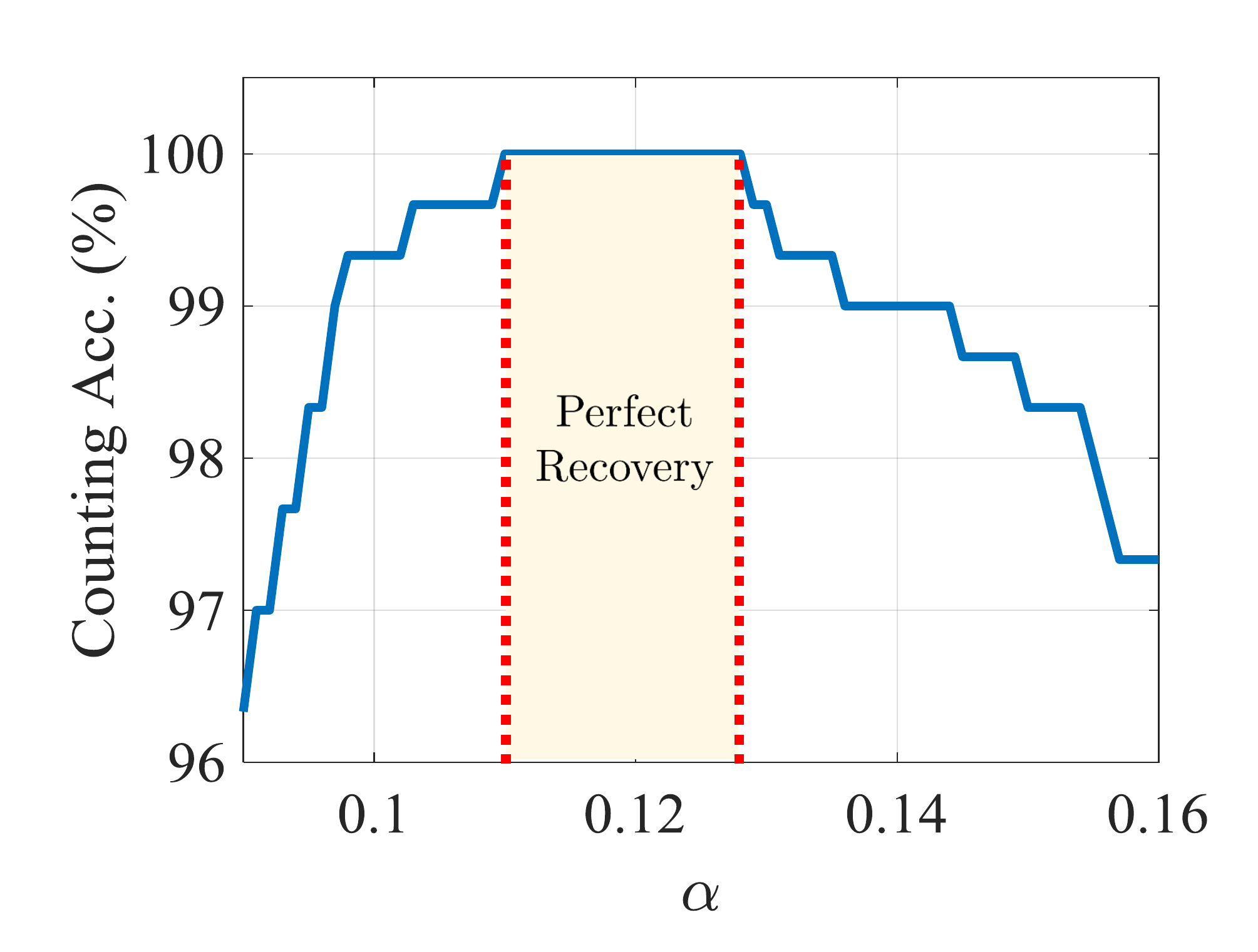}}
\centering
\caption{Counting Accuracy as a function of the threshold parameter $\alpha$.}
\label{fig:count}
\end{figure}

We first examine the ability of the proposed method to estimate the number of speakers in the mixture. Here, we use a smaller frequency range between $0.5-1.5$kHz, which yields better results for the task of counting the number of speakers. We conduct $100$ Monte-Carlo trials for each $J\in\{1,2,3\}$, in which the angles and the distances of the speakers, as well as their input sentences, are randomly selected. Figure~\ref{fig:count} depicts the average counting accuracy as a function of the threshold parameter $\alpha$~\eqref{eq:Jcov} in the range between $0.09$ and $0.16$. We observe that the counting accuracy is robust to the choice of the threshold value, with above $96\%$ accuracy in the defined range. Perfect recovery is obtained for threshold values between $0.11$ and $0.128$.

\begin{figure}[t!]
\centering
\subfigure[]{\includegraphics[width=0.37\textwidth,height=0.2\textheight]{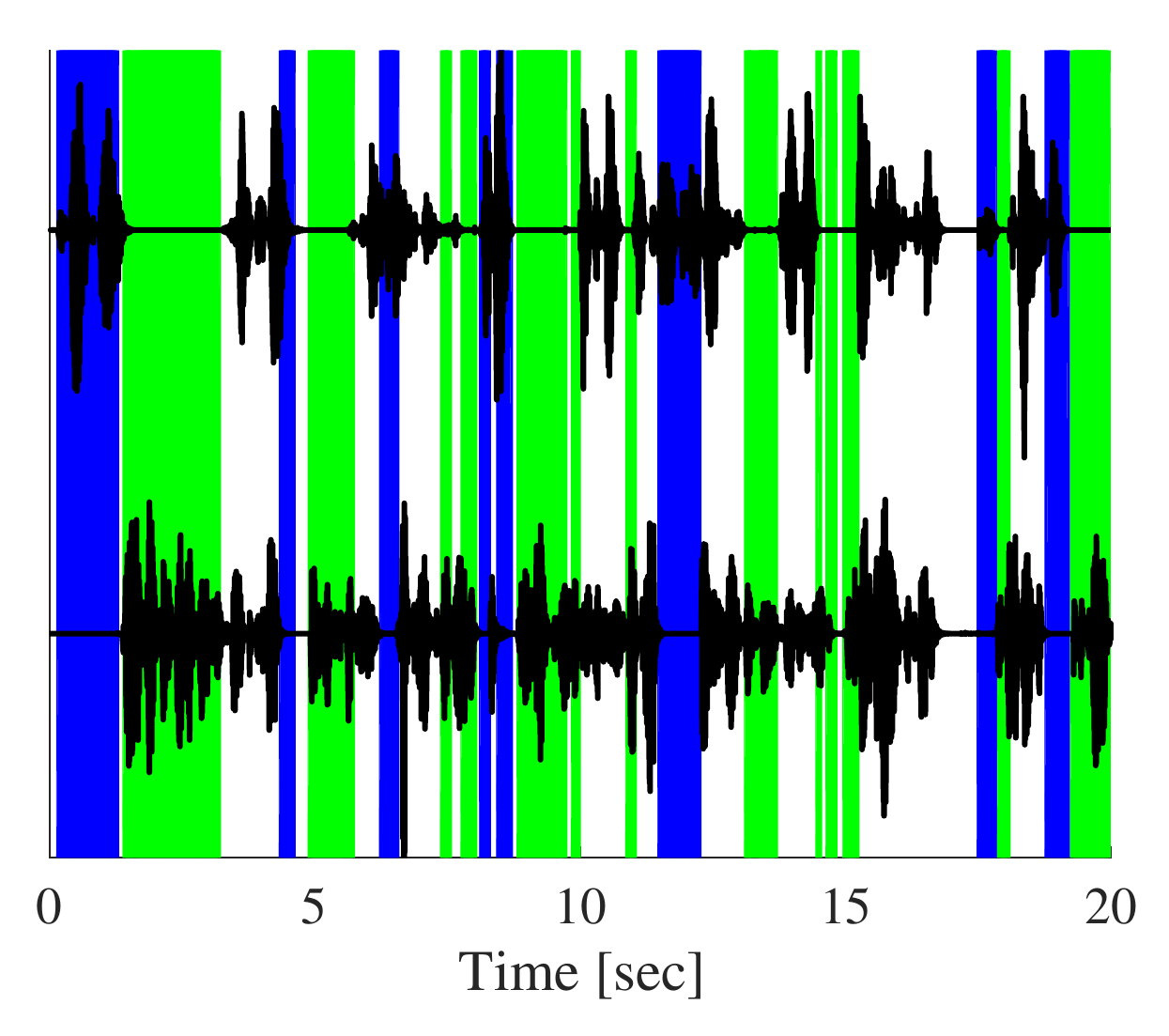}}
\subfigure[]{\includegraphics[width=0.38\textwidth,height=0.23\textheight]{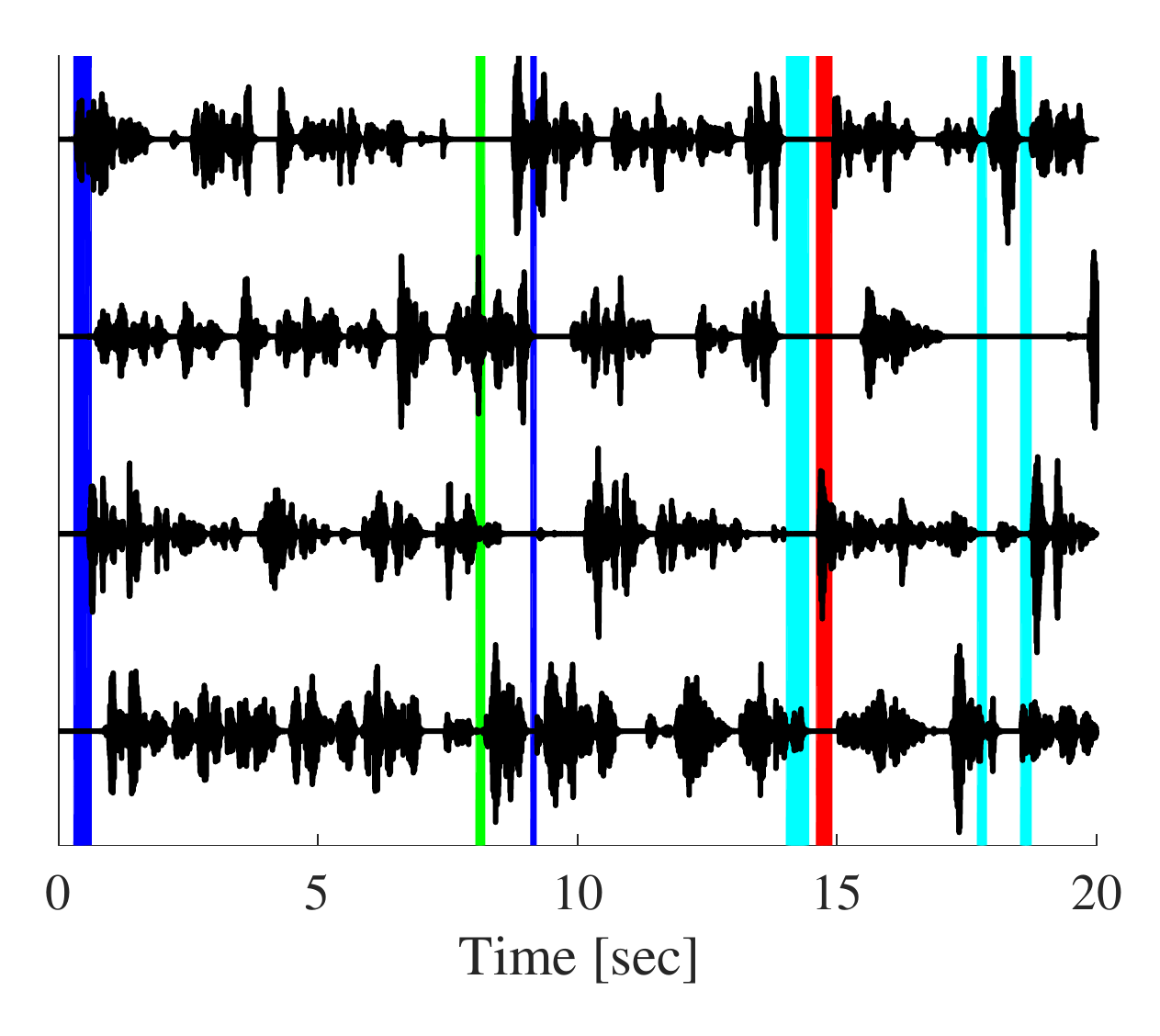}}
\centering
\caption{Time-domain waveforms of each of the speakers for mixtures of (a) $J=2$ and (b) $J=4$ speakers. Time instances, which were detected to be dominated by each of the speakers, are shaded in compatible colors: blue for the first speaker (top), green for the second speaker, red for the third speaker, and cyan for the fourth speaker (down).}
\label{fig:spp}
\end{figure}

Next, we examine the ability of the proposed method to identify the set of frames $\{\mathcal{L}_j\}_{j=1}^J$ dominated by each speaker. Figure~\ref{fig:spp} illustrates the time-domain signals of each of the speakers for a mixture of $J=2$ speakers (a), and for a mixture of $J=4$ speakers (b). The shaded areas stand for time instances which were found to be dominated by each of the speakers, using~\eqref{eq:Lj}. It can be seen that the proposed algorithm successfully identifies time-periods for which one speaker is dominant over the other speakers. Comparing Fig.~\ref{fig:spp}(a) and (b), we observe that as more speakers are involved in the mixture, then less time-periods are dominated by a single speaker.

\begin{table*}[!t]
\centering
\caption{Separation Performance depending on the number of speakers (RT=$360$ms)}
\label{tab:res_J}
\resizebox{0.99 \linewidth}{!}{%
\begin{tabular}{c c c c c c c c c c c c c c c c c c c c c c c }
\toprule
\toprule
&& Input SIR &&& \multicolumn{7}{c}{SIR} &&& \multicolumn{7}{c}{SDR} && \\
\cmidrule{3-3} \cmidrule{6-12} \cmidrule{15-21}

&&&&& \multicolumn{1}{c}{Ideal} && \multicolumn{1}{c}{Semi-Ideal} && \multicolumn{1}{c}{Proposed}&& \multicolumn{1}{c}{NMF} &&&
\multicolumn{1}{c}{Ideal} && \multicolumn{1}{c}{Semi-Ideal} && \multicolumn{1}{c}{Proposed}&& \multicolumn{1}{c}{NMF} && \\

\cmidrule{6-12} \cmidrule{15-21}

\multirow{1}{*}{2 Speakers}
&&$0$dB&&&      21.2 &&  19.5  &&  18.1  &&  14.3  &&&
                8.5 &&  8.4    &&  7.2   &&  8.2 &&  \\

\cmidrule{3-3} \cmidrule{6-12} \cmidrule{15-21}

\multirow{1}{*}{3 Speakers}
&&$-3.2$dB&&&      16.4 &&  13  &&  11.9  &&  9.3  &&&
                  6.3 &&  5.2   &&  4.6   &&  4.6   &&  \\

\cmidrule{3-3} \cmidrule{6-12} \cmidrule{15-21}

\multirow{1}{*}{4 Speakers}
&&$-5$dB&&&      13.3 &&  10.4  &&  9.9  &&  7  &&&
                    6.8 &&  4.8  &&  4.5  &&  2.7  &&  \\
\bottomrule
\bottomrule
\end{tabular}}
\end{table*}

\begin{table*}[!t]
\centering
\caption{Separation Performance depending on reverberation time (3 speakers)}
\label{tab:res_rev}
\resizebox{0.99 \linewidth}{!}{%
\begin{tabular}{c c c c c c c c c c c c c c c c c c c c c c c}
\toprule
\toprule
&& Input SIR &&& \multicolumn{7}{c}{SIR} &&& \multicolumn{7}{c}{SDR} && \\
\cmidrule{3-3} \cmidrule{6-12} \cmidrule{15-21}
&&&&& \multicolumn{1}{c}{Ideal} && \multicolumn{1}{c}{Semi-Ideal} && \multicolumn{1}{c}{Proposed}&& \multicolumn{1}{c}{NMF} &&&
\multicolumn{1}{c}{Ideal} && \multicolumn{1}{c}{Semi-Ideal} && \multicolumn{1}{c}{Proposed}&& \multicolumn{1}{c}{NMF} && \\
\cmidrule{6-12} \cmidrule{15-21}

\multirow{1}{*}{$160$ms}
&&$-3.4$dB&&&      19.1 &&  17.8  &&  17.3  &&  8.6  &&&
                 13.7 &&  12.9  &&  12.3    &&  4.5    &&\\
\cmidrule{3-3} \cmidrule{6-12} \cmidrule{15-21}

\multirow{1}{*}{$360$ms}
&&$-3.2$dB&&&      16.4 &&  13  &&  11.9  &&  9.3  &&&
                  6.3 &&  5.2   &&  4.6   &&  4.6   &&  \\
\cmidrule{3-3} \cmidrule{6-12} \cmidrule{15-21}

\multirow{1}{*}{$610$ms}
&&$-3.2$dB&&&      16.7 &&  13.1  &&  12.3   &&  9.5 &&&
                  4.5 &&  3.6   &&  3.2   &&  3.9   &&  \\
\bottomrule
\bottomrule
\end{tabular}}
\end{table*}

The separation performance is evaluated using the \ac{SIR} and \ac{SDR} measures, evaluated using the \ac{BSS}-Eval toolbox~\cite{vincent2006performance}.
The measures are averaged over $20$ Monte-Carlo trials, in which the angles and the distances of the sources, as well as their input sentences, are randomly selected.

We compare the proposed method to two oracle methods, which are also based on the unmixing scheme of~\eqref{eq:unmix}. In addition, we compare to a multichannel \ac{NMF} algorithm~\cite{ozerov2010multichannel} representing state of-the-art algorithms of the \ac{BSS} family. The methods based on~\eqref{eq:unmix} use either of the following procedures for estimating the \acp{RTF}, used to compute the unmixng matrix:
\begin{enumerate}
\item Ideal: The \acp{RTF} are estimated using the individually measured signals, i.e.:
\begin{equation}
\hat{H}^m_j(k)=\frac{\sum_{l=1}^L Y_j^m(l,f)Y^{1*}_j(l,f)}{\sum_{l=1}^L Y^1_j(l,f)Y^{1*}_j(l,f)}
\label{eq:RTFtrue}
\end{equation}
\item Semi-Ideal: The \acp{RTF} are estimated by~\eqref{eq:RTFest} based on the measured mixtures~\eqref{eq:measured}, where the sets $\{\mathcal{L}_j\}_{j=1}^J$ are determined using the oracle speakers' probabilities computed by:
\begin{equation}
l\in \mathcal{L}_j, \>\>\>  \mbox{if } \>\> \frac{\sum_{k}\|Y^1_j(l,f)\|^2}{\sum_{j=1}^J \sum_{k}\|Y^1_i(l,f)\|^2}>\gamma
\label{eq:spp}
\end{equation}
where $\gamma$ is a threshold set to $0.95$, $0.9$ or $0.8$ for $J=2$, $J=3$ or $J=4$, respectively.
\item Proposed: The \acp{RTF} are estimated by~\eqref{eq:RTFest}, where the sets $\mathcal{L}_j,\>1\leq j \leq J$ are determined using the proposed algorithm, presented in Section~\ref{sec:count_seperate}, where $\beta$ is set to $0.95$.
\end{enumerate}

The parameters of the \ac{NMF} algorithm are initialized using the separated speakers, which are artificially mixed with \ac{SIR} that is improved with respect to the input \ac{SIR} of the given mixture by $3$dB.

We evaluate the performance of all the algorithms depending on the number of speakers and on the reverberation time. The results depending on the number of speakers are depicted in Table~\ref{tab:res_J} for $J=\{2,3,4\}$, with a fixed reverberation time of $360$ms. The results depending on the reverberation time are depicted in Table~\ref{tab:res_rev} for $T_{60}=\{160,360,610\}$ms, for mixtures of $J=3$ speakers.

\begin{figure*}[t!]
\centering
\subfigure[]{\includegraphics[width=0.325\textwidth,height=0.2\textheight]{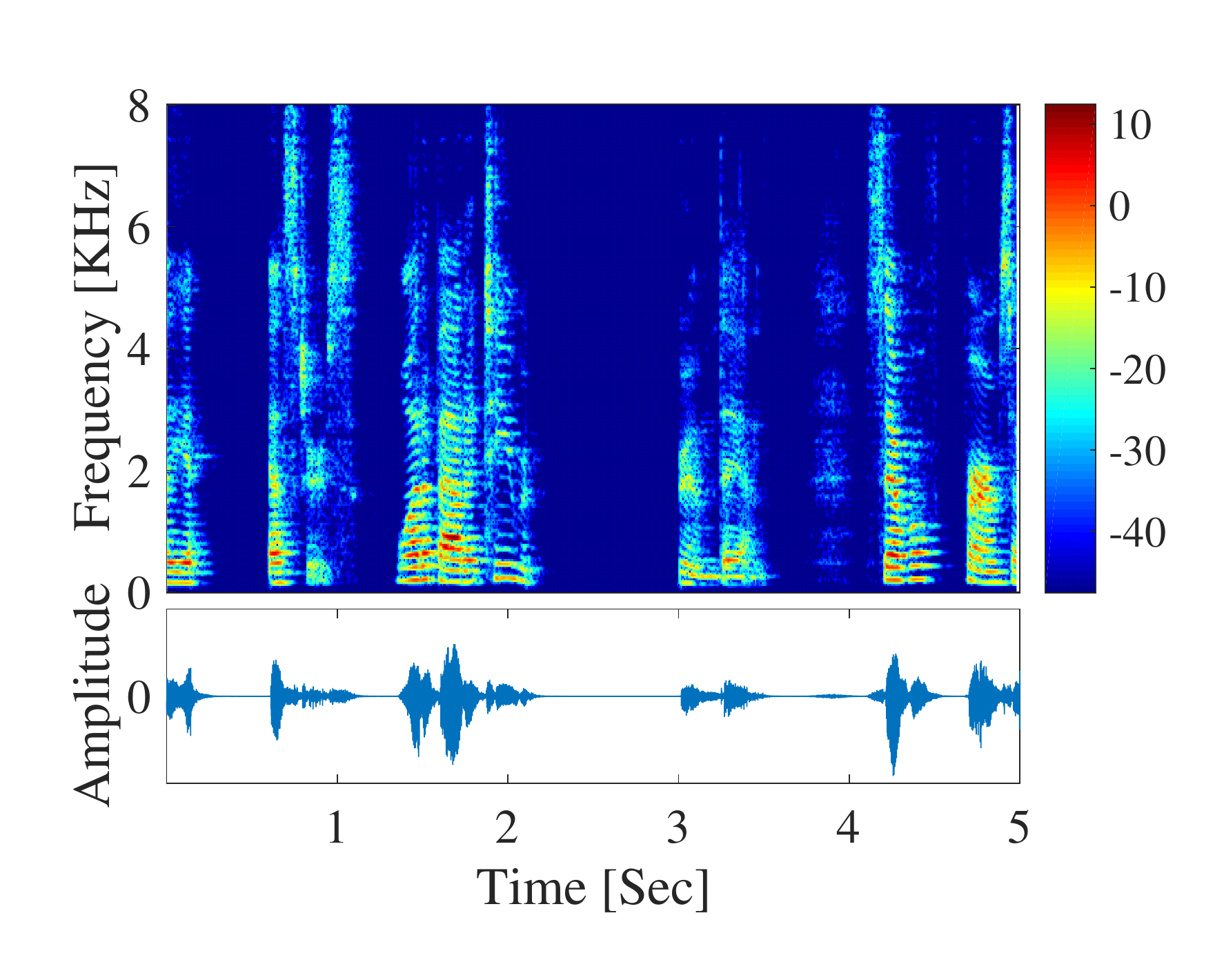}}
\subfigure[]{\includegraphics[width=0.325\textwidth,height=0.2\textheight]{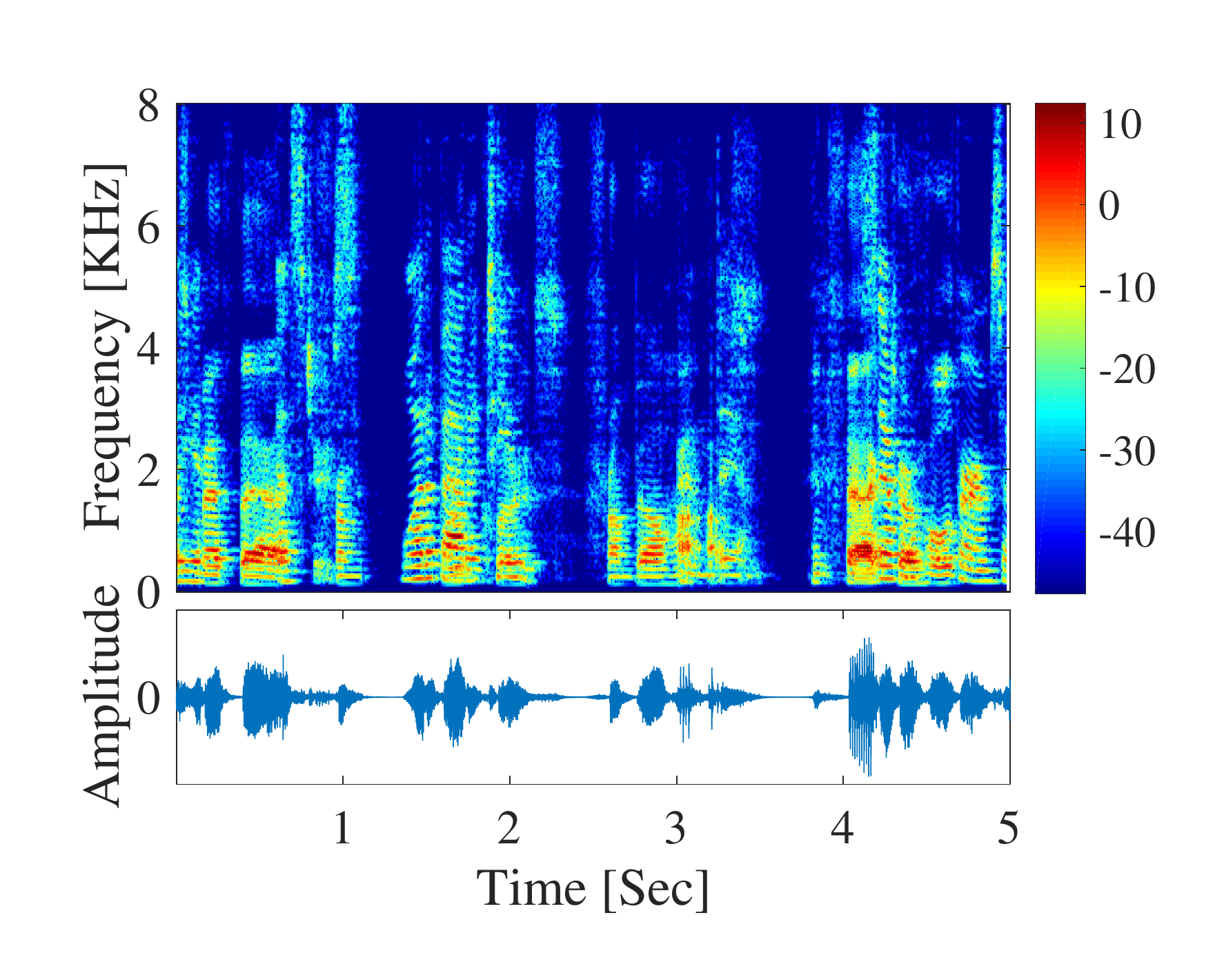}}
\subfigure[]{\includegraphics[width=0.325\textwidth,height=0.2\textheight]{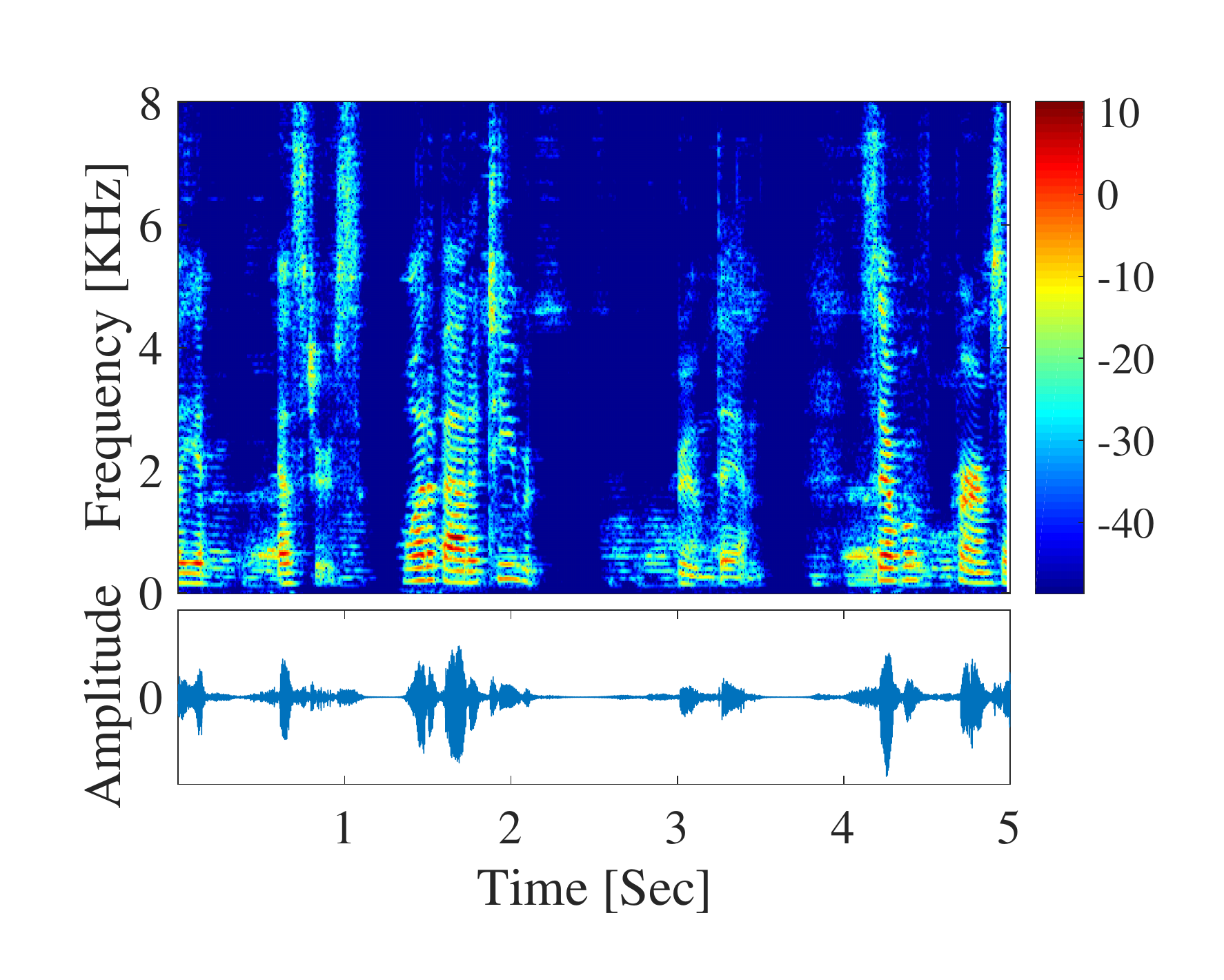}}
\centering
\caption{Spectrograms and waveforms of the first speaker at the first microphone~(a), the two speakers at the first microphone~(b), the estimated first speaker~(c).}
\label{fig:spec}
\end{figure*}

We observe that the ideal unmixing yields the best results. In fact, it represents an upper bound for the separation capabilities, since it is derived using the separated speakers. The semi-ideal unmixing is inferior with respect to the upper bound, since the ideal unmixing uses the original signals for estimating the \acp{RTF}, whereas the semi-ideal unmixing uses non-pure frames from the mixed signals, which may contain also low energy components of other speakers. The proposed estimator determines the frames dominated by a certain speaker based on the mixed signals. Its performance is comparable to the semi-ideal unmixing with a small gap of $0.3-1.5$dB. The \ac{NMF} method is inferior with respect to the proposed method in almost all cases. It should be emphasized that the \ac{NMF} algorithm uses an initialization with improved \ac{SIR}, whereas the proposed method is completely blind. For all algorithms, a performance degradation is observed as the number of speakers increases or as the reverberation time increases. It should be noted that for both the semi-ideal unmixing and the proposed method, an increase in the number of speakers means a decrease in the number of frames dominated by a single speaker, hence, the performance gap between both algorithms and the ideal unmixing increases.

Figure~\ref{fig:spec} presents an example of the spectrograms and the waveforms of a mixture of $J=2$ speakers, where the first speaker~(a), the mixture~(b), and the output signal of the proposed method~(c), are depicted. It is evident that the spectral components of the second speaker are significantly attenuated, while preserving most of the spectral components of the first speaker. There is also a good match between the original and the output waveforms.

\section{Conclusions}
\label{sec:conc}
We present a novel framework for speech source counting and separation in a completely blind manner. The separation is based on the sparsity of speech in the \ac{STFT} domain, as well as the fact that each speaker is associated with a unique spatial signature, manifested by the \ac{RTF} between the speaker and the microphones. A spectral decomposition of the correlation matrix of different time frames reveals the number of speakers, and forms a simplex of the speakers' probabilities across time. Utilizing convex geometry tools, the frames dominated by each speaker are identified. The \acp{RTF} of the different speakers are estimated using these identified frames, and an unmixing scheme is implemented to separate the individual speakers. The performance is demonstrated in an experimental study for various reverberation levels.

\begin{appendices}
\section{}
\label{sec:appA}
In this section, we compute the expected correlation between observations and evaluate its variance. The computation is based on the statistical model of Section~\ref{sec:stat}. Recall the following assumption regarding the hidden sources:
\begin{equation}
E\left\{h_i(k)h_j(\tilde{k})\right\}=\delta_{ij}\cdot\delta_{k\tilde{k}}.
\label{eq:comp}
\end{equation}
which follows from~\eqref{eq:dist}, the zero-mean assumption and the mutual independence of the hidden sources.
In addition, the indicator functions satisfy (recall ~\eqref{eq:indi0}):
\begin{equation}
I_j(l,k)I_i(l,k)=I_j(l,k)\delta_{ij}.
\label{eq:indi}
\end{equation}

We compute the correlation for $1\leq l,n \leq L,l\neq n$:
\begin{align*}
&E\left\{\frac{1}{D}\mathbf{a}^T(l)\mathbf{a}(n)\right\}\\ \numberthis \label{eq:corrln}
&=\frac{1}{D}E\left\{\sum_{k=1}^D\sum_{i,j=1}^JI_i(l,k)I_j(n,k)h_i(k)h_j(k)\right\}\\ \nonumber
&=\frac{1}{D}\sum_{k=1}^D\sum_{i,j=1}^JE\left\{I_i(l,k)I_j(n,k)\right\}E\left\{h_i(k)h_j(k)\right\}\\ \nonumber
&=\frac{1}{D}\sum_{k=1}^D\sum_{i,j=1}^JE\left\{I_i(l,k)\right\}E\left\{I_j(n,k)\right\}E\left\{h_i(k)h_j(k)\right\}\\ \nonumber
&=\frac{1}{D}\sum_{k=1}^D\sum_{i,j=1}^Jp_i(l)p_j(n)\delta_{ij}\\ \nonumber
&=\sum_{j=1}^Jp_j(l)p_j(n).\\ \nonumber
\end{align*}
The second equality follows from the independence of the indicator functions and the sources. The third equality follows from the independence of the indicator functions for $l\neq n$. The fourth equality is due to~\eqref{eq:comp}.

For $l=n$ the autocorrelation is given by:
 \begin{align*}
&E\left\{\frac{1}{D}\mathbf{a}^T(l)\mathbf{a}(l)\right\}\\ \numberthis \label{eq:corrll}
&=\frac{1}{D}E\left\{\sum_{k=1}^D\sum_{i,j=1}^JI_i(l,k)I_j(l,k)h_i(k)h_j(k)\right\}\\ \nonumber
&=\frac{1}{D}\sum_{k=1}^D\sum_{i,j=1}^JE\left\{I_i(l,k)I_j(l,k)\right\}E\left\{h_i(k)h_j(k)\right\}\\ \nonumber
&=\frac{1}{D}\sum_{k=1}^D\sum_{j=1}^J E\left\{I_j(l,k)\right\}E\left\{h^2_j(k)\right\}\\ \nonumber
&=\sum_{j=1}^Jp_j(l)=1\\ \nonumber
\end{align*}
where the third equality follows from~\eqref{eq:indi}.

We compute the variance of~\eqref{eq:corrln}:
\begin{align*}
&\textrm{Var}\left\{\frac{1}{D}\mathbf{a}(l)^T\mathbf{a}(n)\right\}\\ \numberthis \label{eq:var}
&=\frac{1}{D^2}E\left\{(\mathbf{a}(l)^T\mathbf{a}(n))^2\right\}-\frac{1}{D^2}E^2\left\{\mathbf{a}(l)^T\mathbf{a}(n)\right\}.\\\nonumber
\end{align*}
We show that the variance~\eqref{eq:var} approaches zero for $D$ large enough, implying that the typical value $\frac{1}{D}\mathbf{a}^T(l)\mathbf{a}(n)$ approaches the expected value $\frac{1}{D}E\left\{\mathbf{a}^T(l)\mathbf{a}(n)\right\}$.

The first moment is given in~\eqref{eq:corrln}. We compute the second moment for $1\leq l,n \leq L,l\neq n$:
\begin{align*}
&E\left\{(\mathbf{a}^T(l)\mathbf{a}(n))^2\right\}\\ \numberthis \label{eq:corrln2}
&=E\left\{\left(\sum_{k=1}^D\sum_{i,j=1}^JI_i(l,k)I_j(n,k)h_i(k)h_j(k)\right)^2\right\}\\ \nonumber
&=\sum_{k,\tilde{k}=1}^D\sum_{i,j,\tilde{i},\tilde{j}=1}^JE\left\{I_i(l,k)I_j(n,k)I_{\tilde{i}}(l,\tilde{k})I_{\tilde{j}}(n,\tilde{k})\right\} \\\nonumber
&\cdot E\left\{h_i(k)h_j(k)h_{\tilde{i}}(\tilde{k})h_{\tilde{j}}(\tilde{k})\right\}.\\ \nonumber
\end{align*}
Splitting the sum over $\tilde{k}$ into two parts, for $\tilde{k}=k$ and for $\tilde{k}\neq k$, we receive:
\begin{align*}
&E\left\{(\mathbf{a}^T(l)\mathbf{a}(n))^2\right\}\\ \numberthis \label{eq:corrln2a}
&=\sum_{k=1}^D\sum_{i,j,\tilde{i},\tilde{j}=1}^JE\left\{I_i(l,k)I_j(n,k)I_{\tilde{i}}(l,k)I_{\tilde{j}}(n,k)\right\} \\\nonumber
&\cdot E\left\{h_i(k)h_j(k)h_{\tilde{i}}(k)h_{\tilde{j}}(k)\right\}\\ \nonumber
&+\sum_{\substack{k,\tilde{k}=1\\ \tilde{k}\neq k}}^D\sum_{i,j,\tilde{i},\tilde{j}=1}^JE\left\{I_i(l,k)I_j(n,k)I_{\tilde{i}}(l,\tilde{k})I_{\tilde{j}}(n,\tilde{k})\right\} \\\nonumber
&\cdot E\left\{h_i(k)h_j(k)h_{\tilde{i}}(\tilde{k})h_{\tilde{j}}(\tilde{k})\right\}\\ \nonumber
&=\sum_{k=1}^D\sum_{i,j=1}^JE\left\{I_i(l,k)\right\}E\left\{I_j(n,k)\right\}E\left\{h^2_i(k)h^2_j(k)\right\}\\ \nonumber
&+\sum_{\substack{k,\tilde{k}=1\\ \tilde{k}\neq k}}^D\sum_{i,j,\tilde{i},\tilde{j}=1}^JE\left\{I_i(l,k)\right\}E\left\{I_j(n,k)\right\} \\ \nonumber
&E\left\{I_{\tilde{i}}(l,\tilde{k})\right\} E\left\{I_{\tilde{j}}(n,\tilde{k})\right\}
E\left\{h_i(k)h_j(k)h_{\tilde{i}}(\tilde{k})h_{\tilde{j}}(\tilde{k})\right\}\\ \nonumber
\end{align*}
where the second equality follows from~\eqref{eq:indi}, and the independence of the indicator functions for $l\neq n$ or $k \neq \tilde{k}$.
Evaluating the expectations of the indicators, we get:
\begin{align*}
E\left\{(\mathbf{a}(l)^T\mathbf{a}(n))^2\right\}
&=\sum_{k=1}^D\sum_{i,j=1}^Jp_i(l)p_j(n)E\left\{h^2_i(k)h^2_j(k)\right\}\\ \nonumber
&+\sum_{\substack{k,\tilde{k}=1\\ \tilde{k}\neq k}}^D\sum_{i,j,\tilde{i},\tilde{j}=1}^Jp_i(l)p_j(n)p_{\tilde{i}}(l)p_{\tilde{j}}(n)\\ \nonumber
&\cdot E\left\{h_i(k)h_j(k)h_{\tilde{i}}(\tilde{k})h_{\tilde{j}}(\tilde{k})\right\}.\numberthis \label{eq:corrln2b}
\end{align*}
Focusing on the second term in~\eqref{eq:corrln2b}, we further simplify:
\begin{align*}
&\sum_{\substack{k,\tilde{k}=1\\ \tilde{k}\neq k}}^D\sum_{i,j,\tilde{i},\tilde{j}=1}^Jp_i(l)p_j(n)p_{\tilde{i}}(l)p_{\tilde{j}}(n) \\ \numberthis \label{eq:corrln2c1}
&\cdot E\left\{h_i(k)h_j(k)h_{\tilde{i}}(\tilde{k})h_{\tilde{j}}(\tilde{k})\right\}\\ \nonumber
&=\sum_{\substack{k,\tilde{k}=1\\ \tilde{k}\neq k}}^D
\sum_{i,j,\tilde{i},\tilde{j}=1}^Jp_i(l)p_j(n)p_{\tilde{i}}(l)p_{\tilde{j}}(n)\\ \nonumber
&E\Big\{h_i(k)h_j(k)\Big\}
E\Big\{h_{\tilde{i}}(\tilde{k})h_{\tilde{j}}(\tilde{k})\Big\}\\ \nonumber
\end{align*}
where we relied on the independence between $h_i(k)$ and $h_{\tilde{i}}(\tilde{k})$ for $k\neq\tilde{k}$, assumed in~\eqref{eq:comp}. Further relying on the statistical model of~\eqref{eq:comp}, we receive:
\begin{align*}
&\sum_{\substack{k,\tilde{k}=1\\ \tilde{k}\neq k}}^D
\sum_{i,j,\tilde{i},\tilde{j}=1}^Jp_i(l)p_j(n)p_{\tilde{i}}(l)p_{\tilde{j}}(n)\\ \numberthis \label{eq:corrln2c2}
&E\Big\{h_i(k)h_j(k)\Big\}
E\Big\{h_{\tilde{i}}(\tilde{k})h_{\tilde{j}}(\tilde{k})\Big\}\\ \nonumber
&=\sum_{\substack{k,\tilde{k}=1\\ \tilde{k}\neq k}}^D
\sum_{i,j,\tilde{i},\tilde{j}=1}^Jp_i(l)p_j(n)p_{\tilde{i}}(l)p_{\tilde{j}}(n)\delta_{ij}\delta_{\tilde{i}\tilde{j}}\\ \nonumber
&=D(D-1)\sum_{j,\tilde{j}=1}^Jp_j(l)p_j(n)p_{\tilde{j}}(l)p_{\tilde{j}}(n)\\ \nonumber
&=D(D-1)\left(\sum_{j=1}^Jp_j(l)p_j(n)\right)^2
\end{align*}
Substituting~\eqref{eq:corrln2c2} into~\eqref{eq:corrln2b}, we get:
\begin{align*}
&E\left\{(\mathbf{a}^T(l)\mathbf{a}(n))^2\right\}\\ \numberthis \label{eq:corrln2d}
&=\sum_{k=1}^D\sum_{i,j=1}^Jp_i(l)p_j(n)E\left\{h^2_i(k)h^2_j(k)\right\}\\ \nonumber
&+D(D-1)\left(\sum_{j=1}^Jp_j(l)p_j(n)\right)^2
\end{align*}
Substituting~\eqref{eq:corrln} and~\eqref{eq:corrln2d} into~\eqref{eq:var}, we receive:
\begin{align*}
&\textrm{Var}\left\{\frac{1}{D}\mathbf{a}(l)^T\mathbf{a}(n)\right\} \\ \numberthis \label{eq:corrln2e}
&=\frac{1}{D^2}E\left\{(\mathbf{a}(l)^T\mathbf{a}(n))^2\right\}-\frac{1}{D^2}E^2\left\{\mathbf{a}(l)^T\mathbf{a}(n)\right\} \\
&=\frac{1}{D^2}\sum_{k=1}^D\sum_{i,j=1}^Jp_i(l)p_j(n)E\left\{h^2_i(k)h^2_j(k)\right\}\\ \nonumber
&+\frac{D(D-1)}{D^2}\left(\sum_{j=1}^Jp_j(l)p_j(n)\right)^2-\left(\sum_{j=1}^Jp_j(l)p_j(n)\right)^2\\ \nonumber
\end{align*}
For $D$ large enough, we have $\frac{D-1}{D}\approx 1$, and~\eqref{eq:corrln2e} simplifies to:
\begin{align*}
&\textrm{Var}\left\{\frac{1}{D}\mathbf{a}(l)^T\mathbf{a}(n)\right\} \\ \numberthis \label{eq:corrln2f}
&\approx\frac{1}{D^2}\sum_{k=1}^D\sum_{i,j=1}^Jp_i(l)p_j(n)E\left\{h^2_i(k)h^2_j(k)\right\}\\ \nonumber
&=\frac{1}{D^2}\sum_{k=1}^D\sum_{j=1}^Jp_j(l)p_j(n)E\left\{h_j^4(k)\right\}\\ \nonumber
&+\frac{1}{D^2}\sum_{k=1}^D\sum_{\substack{i,j=1\\ i\neq j}}^Jp_i(l)p_j(n)E\left\{h^2_i(k)\right\}E\left\{h^2_j(k)\right\}\\ \nonumber
&=\frac{1}{D^2}\sum_{k=1}^D\sum_{j=1}^Jp_j(l)p_j(n)E\left\{h_j^4(k)\right\}
+\frac{1}{D}\sum_{\substack{i,j=1\\ i\neq j}}^Jp_i(l)p_j(n).
\end{align*}
In the second term in~\eqref{eq:corrln2f}, we have:
\begin{align*}
\sum_{j=1}^J\left(p_j(n)\sum_{\substack{i=1\\ i\neq j}}^Jp_i(l)\right)
&=\sum_{j=1}^Jp_j(n)\left(1-p_j(l)\right) \\ \numberthis \label{eq:corrln2g}
&=1-\sum_{j=1}^Jp_j(l)p_j(n)
\end{align*}
Let $E\left\{h_j^4(k)\right\}\equiv C_{4}$, substituting~\eqref{eq:corrln2g} into~\eqref{eq:corrln2f}, we get:
\begin{align*}
&\textrm{Var}\left\{\frac{1}{D}\mathbf{a}(l)^T\mathbf{a}(n)\right\} \\ \numberthis \label{eq:corrln2h}
&\approx\frac{C_4}{D}\sum_{j=1}^Jp_j(l)p_j(n)+\frac{1}{D}\left(1-\sum_{j=1}^Jp_j(l)p_j(n)\right). \\ \nonumber
&=\frac{C_4-1}{D}\sum_{j=1}^Jp_j(l)p_j(n)+\frac{1}{D}
\leq  \frac{C_4-1}{D}+ \frac{1}{D} =\frac{C_4}{D}.
\end{align*}
For zero-mean Gaussian sources $C4=E\left\{h_j^4(k)\right\}=3\cdot E\left\{h_j^2(k)\right\}$, which under the unit variance  assumption amounts to $C4=3$. Hence, we can easily set the value of $D$, satisfying $D\gg C_4$. Accordingly, we get:
\begin{equation}
\textrm{Var}\left\{\frac{1}{D}\mathbf{a}^T(l)\mathbf{a}(n)\right\} \approx 0.
\end{equation}
We conclude that for $D$ large enough the typical value of $\frac{1}{D}\mathbf{a}^T(l)\mathbf{a}(n)$ is close to its expected value $\frac{1}{D}E\left\{\mathbf{a}^T(l)\mathbf{a}(n)\right\}$. Hence, $\frac{1}{D}\mathbf{a}^T(l)\mathbf{a}(n)$ can be used instead of its expected value.

\section{}
\label{sec:appB}
In this section, we discuss the spectral decomposition of the correlation matrix $\mathbf{W}$, and its approximation as $\mathbf{W}\approx\mathbf{PP}^T$. Recall the following representation of the correlation matrix $\mathbf{W}$ (Eq.~\eqref{eq:PP}):
\begin{equation}
\mathbf{W}=\mathbf{P}\mathbf{P}^T + \Delta \mathbf{W}
\label{eq:PP2}
\end{equation}
where $\Delta \mathbf{W}$ is a diagonal matrix with $\Delta W_{ll}=1-\sum_{j=1}^Jp^2_{j}(l)$. Here, we analyse the influence of $\Delta \mathbf{W}$ on the obtained spectral decomposition, and show that it has a negligible affect on the proposed speaker counting and separation method.

For this purpose, we use matrix perturbation theory~\cite{stewart1990matrix}. Consider the perturbed matrix $\mathbf{W}$ given by:
\begin{equation}
\mathbf{W}=\mathbf{K}+\Delta\mathbf{W}
\end{equation}
where the matrix $\Delta\mathbf{W}$ represents a small perturbation. According to the matrix perturbation theory~\cite{stewart1990matrix}, the following Theorem relates the \acp{EVD} of the matrices $\mathbf{W}$ and $\mathbf{K}$:
\begin{theorem}
Let $\{\lambda_j,\mathbf{u}_j\}_j$ be the set of eigenvalues and eigenvectors of the matrix $\mathbf{K}$, and let $\{\tilde{\lambda}_j,\tilde{\mathbf{u}}_j\}_j$ be the set of eigenvalues and eigenvectors of the matrix $\mathbf{W}=\mathbf{K}+\Delta\mathbf{W}$. Then:
\begin{align}
\tilde{\lambda}_j&=\lambda_j+\mathbf{u}^T_j\Delta\mathbf{W}\mathbf{u}_j+O\left(\|\Delta\mathbf{W}\|^2\right)  \label{eq:pertl} \\
\tilde{\mathbf{u}}_j&=\mathbf{u}_j+\sum_{i\neq j}\frac{\mathbf{u}^T_i\Delta\mathbf{W}\mathbf{u}_j}{\lambda_j-\lambda_i}\mathbf{u}_i+O\left(\|\Delta\mathbf{W}\|^2\right)
\numberthis \label{eq:pertu}
\end{align}
\label{th1}
\end{theorem}
According to Theorem~\ref{th1}, each eigenvalue $\tilde{\lambda}_j$ of the perturbed matrix deviates from the corresponding eigenvalue $\lambda_j$ of the original matrix by the weighted norm $\mathbf{u}^T_j\Delta\mathbf{W}\mathbf{u}_j$. In addition, each perturbed eigenvector $\tilde{\mathbf{u}}_j$ equals the corresponding original eigenvector $\mathbf{u}_j$  plus a term, which consists of the contributions of the other eigenvectors of the original matrix. The contribution of the other eigenvectors is proportional to the weighted inner product $\mathbf{u}^T_i\Delta\mathbf{W}\mathbf{u}_j$ divided by the difference $\lambda_j-\lambda_i$ between the corresponding eigenvalues.

In our case, the original matrix $\mathbf{K}\equiv\mathbf{PP}^T$ has a rank-J decomposition. Accordingly, $\mathbf{PP}^T$ has $J$ nonzero eigenvalues $\Lambda_{1}\equiv\{\lambda_j\}_{j=1}^{J}$, associated with $J$ eigenvectors $\mathcal{U}_{1}\equiv\{\mathbf{u}_j\}_{j=1}^{J}$ that span the column space of the matrix $\mathbf{P}$. In addition, there are $L-J$ zero eigenvalues $\Lambda_{0}\equiv\{\lambda_j\}_{j=J+1}^{L}$, associated with $L-J$ eigenvectors $\mathcal{U}_{0}\equiv\{\mathbf{u}_j\}_{j=J+1}^{L}$ that span the null space of $\mathbf{P}$.

The weighted inner product can be written as:
\begin{equation}
\mathbf{u}^T_i\Delta\mathbf{W}\mathbf{u}_j=\mathbf{v}^T_i\mathbf{v}_j=\|\mathbf{v}_i\|\|\mathbf{v}_j\|\cos\theta_{ij}
\label{eq:product}
\end{equation}
where $\mathbf{v}_j=\mathbf{A}\mathbf{u}_j$ with $\Delta\mathbf{W}=\mathbf{A}^T\mathbf{A}$, and $\theta_{ij}$ is the angle between $\mathbf{v}_i$ and $\mathbf{v}_j$. In our case, $\mathbf{A}$ is a diagonal matrix with elements $A_{ll}=\sqrt{1-\sum_{j=1}^Jp_j^2(l)}\leq1$, implying $\|\mathbf{v}_j\|\leq\|\mathbf{u}_j\|=1$. We assume that multiplication  by $\mathbf{A}$ only slightly affect the right angle between the orthonormal vectors $\mathbf{u}_i$ and $\mathbf{u}_j$ for $i\neq j$, implying $\cos\theta_{ij}\approx \epsilon$. Hence, we get the following bound:
\begin{equation}
\left|\mathbf{u}^T_i\Delta\mathbf{W}\mathbf{u}_j\right| \leq
\left\{
	\begin{array}{ll}
		\epsilon &  \mbox{ if } i\neq j\\
		1 &    \mbox{ if } i = j
	\end{array}
\right..
\label{eq:bound}
\end{equation}

Accordingly, the eigenvalue perturbation is limited to $1$ and the eigenvector perturbation depends on the ratio $\frac{\epsilon}{\lambda_j-\lambda_i}$. An eigenvector $\mathbf{u}_{j^*}$ will have small contribution from the eigenvector $\mathbf{u}_{i}$, when  $|\lambda_{j*}-\lambda_i|\gg \epsilon$.

Note that in the proposed algorithm we are only interested in the eigenvectors in $\mathcal{U}_{1}$, spanning the column space of $\mathbf{PP}^T$. For a particular $\mathbf{u}_{j^*}\in \mathcal{U}_{1}$, there may be some contribution of the other eigenvectors in $\mathcal{U}_{1}$, depending on the respective eigenvalues decay. The contribution of eigenvectors in $\mathcal{U}_{0}$, associated with zero eigenvalues, is necessarily smaller and is negligible for $|\lambda_{j*}|\gg \epsilon$.

We demonstrate the conclusions of the above analysis using the example of Section~\ref{sec:corr_anl}.
We compute the eigenvectors of $\mathbf{PP}^T$ and of $\mathbf{W}$, and measure their correlation for $J=3$. We present the correlation between the first $3$ eigenvectors of $\mathbf{W}$ and the first $5$ eigenvectors of $\mathbf{PP}^T$:
\begin{equation}
\begin{bmatrix}
1      &   -4e^{-16}   & -6e^{-5} & -6e^{-5} &  -3e^{-5}\\
4e^{-5}   &   1        & -5e^{-3} & -4e^{-6} &  2e^{-5}\\
-6e^{-5}   &   -5e^{-3}        & 1 & 1e^{-4} &  -7e^{-5}\\
\end{bmatrix}.
\nonumber
\end{equation}
where the $(i,j)$th element equals $\tilde{\mathbf{u}}^T_i\mathbf{u}_j$.
We deduce that $\tilde{\mathbf{u}}_j\approx \mathbf{u}_j$ for $1\leq j \leq 3$, i.e. the first $J$ eigenvectors of $\mathbf{W}$ are almost identical to the first $J$ eigenvectors of $\mathbf{PP}^T$. We also compare between the first $5$ eigenvalues of both matrices:
\begin{align*}
\lambda_1&=167,\> \lambda_2=44,\> \lambda_3=37,\> \lambda_4=8e^{-15},\> \lambda_5=8e^{-15}\\ \nonumber
\tilde{\lambda}_1&=168,\> \tilde{\lambda}_2=44,\> \tilde{\lambda}_3=38,\> \tilde{\lambda}_4=0.7,\> \tilde{\lambda}_5=0.7. \numberthis \label{eq:eig1}
\end{align*}
We observe that $|\tilde{\lambda}_j-\lambda_j|<1$ as expected. Note that the slight differences between the eigenvalues, seem to have a minor impact on the decision rule of~\eqref{eq:Jcov}, for counting the number of sources. We conclude that the derivations in Section~\ref{sec:stat}, regarding the spectral decomposition of the matrix $\mathbf{PP}^T$, apply also for the correlation matrix $\mathbf{W}$.

\end{appendices}
\balance
\bibliographystyle{IEEEtran}

\begin{thebibliography}{10}
\providecommand{\url}[1]{#1}
\csname url@samestyle\endcsname
\providecommand{\newblock}{\relax}
\providecommand{\bibinfo}[2]{#2}
\providecommand{\BIBentrySTDinterwordspacing}{\spaceskip=0pt\relax}
\providecommand{\BIBentryALTinterwordstretchfactor}{4}
\providecommand{\BIBentryALTinterwordspacing}{\spaceskip=\fontdimen2\font plus
\BIBentryALTinterwordstretchfactor\fontdimen3\font minus
  \fontdimen4\font\relax}
\providecommand{\BIBforeignlanguage}[2]{{%
\expandafter\ifx\csname l@#1\endcsname\relax
\typeout{** WARNING: IEEEtran.bst: No hyphenation pattern has been}%
\typeout{** loaded for the language `#1'. Using the pattern for}%
\typeout{** the default language instead.}%
\else
\language=\csname l@#1\endcsname
\fi
#2}}
\providecommand{\BIBdecl}{\relax}
\BIBdecl

\bibitem{comon2010handbook}
P.~Comon and C.~Jutten, \emph{Handbook of Blind Source Separation: Independent
  component analysis and applications}.\hskip 1em plus 0.5em minus 0.4em\relax
  Academic press, 2010.

\bibitem{lee1998independent}
T.-W. Lee, ``Independent component analysis,'' in \emph{Independent Component
  Analysis}.\hskip 1em plus 0.5em minus 0.4em\relax Springer, 1998, pp. 27--66.

\bibitem{hyvarinen2000independent}
A.~Hyv{\"a}rinen and E.~Oja, ``Independent component analysis: algorithms and
  applications,'' \emph{Neural networks}, vol.~13, no.~4, pp. 411--430, 2000.

\bibitem{hyvarinen2004independent}
A.~Hyv{\"a}rinen, J.~Karhunen, and E.~Oja, \emph{Independent component
  analysis}.\hskip 1em plus 0.5em minus 0.4em\relax John Wiley \& Sons, 2004,
  vol.~46.

\bibitem{cichocki2009nonnegative}
A.~Cichocki, R.~Zdunek, A.~H. Phan, and S.-i. Amari, \emph{Nonnegative matrix
  and tensor factorizations: applications to exploratory multi-way data
  analysis and blind source separation}.\hskip 1em plus 0.5em minus 0.4em\relax
  John Wiley \& Sons, 2009.

\bibitem{zibulevsky2001blind}
M.~Zibulevsky and B.~A. Pearlmutter, ``Blind source separation by sparse
  decomposition in a signal dictionary,'' \emph{Neural computation}, vol.~13,
  no.~4, pp. 863--882, 2001.

\bibitem{makino2007blind}
S.~Makino, T.-W. Lee, and H.~Sawada, \emph{Blind speech separation}.\hskip 1em
  plus 0.5em minus 0.4em\relax Springer, 2007, vol. 615.

\bibitem{pedersen2008convolutive}
M.~S. Pedersen, J.~Larsen, U.~Kjems, and L.~C. Parra, ``Convolutive blind
  source separation methods,'' in \emph{Springer Handbook of Speech
  Processing}.\hskip 1em plus 0.5em minus 0.4em\relax Springer, 2008, pp.
  1065--1094.

\bibitem{vincent2010probabilistic}
E.~Vincent, M.~G. Jafari, S.~A. Abdallah, M.~D. Plumbley, and M.~E. Davies,
  ``Probabilistic modeling paradigms for audio source separation,''
  \emph{Machine Audition: Principles, Algorithms and Systems}, pp. 162--185,
  2010.

\bibitem{mitianoudis2003audio}
N.~Mitianoudis and M.~E. Davies, ``Audio source separation of convolutive
  mixtures,'' \emph{IEEE Transactions on Speech and Audio Processing}, vol.~11,
  no.~5, pp. 489--497, 2003.

\bibitem{sawada2004robust}
H.~Sawada, R.~Mukai, S.~Araki, and S.~Makino, ``A robust and precise method for
  solving the permutation problem of frequency-domain blind source
  separation,'' \emph{IEEE Transactions on Speech and Audio Processing},
  vol.~12, no.~5, pp. 530--538, 2004.

\bibitem{yilmaz2004blind}
O.~Yilmaz and S.~Rickard, ``Blind separation of speech mixtures via
  time-frequency masking,'' \emph{IEEE Transactions on Signal Processing},
  vol.~52, no.~7, pp. 1830--1847, 2004.

\bibitem{fevotte2009nonnegative}
C.~F{\'e}votte, N.~Bertin, and J.-L. Durrieu, ``Nonnegative matrix
  factorization with the itakura-saito divergence: With application to music
  analysis,'' \emph{Neural computation}, vol.~21, no.~3, pp. 793--830, 2009.

\bibitem{ozerov2010multichannel}
A.~Ozerov and C.~F{\'e}votte, ``Multichannel nonnegative matrix factorization
  in convolutive mixtures for audio source separation,'' \emph{IEEE
  Transactions on Audio, Speech, and Language Processing}, vol.~18, no.~3, pp.
  550--563, 2010.

\bibitem{arberet2010robust}
S.~Arberet, R.~Gribonval, and F.~Bimbot, ``A robust method to count and locate
  audio sources in a multichannel underdetermined mixture,'' \emph{IEEE
  Transactions on Signal Processing}, vol.~58, no.~1, pp. 121--133, 2010.

\bibitem{mandel2010model}
M.~I. Mandel, R.~J. Weiss, and D.~P. Ellis, ``Model-based
  expectation-maximization source separation and localization,'' \emph{IEEE
  Transactions on Audio, Speech, and Language Processing}, vol.~18, no.~2, pp.
  382--394, 2010.

\bibitem{traa2014multichannel}
J.~Traa and P.~Smaragdis, ``Multichannel source separation and tracking with
  {RANSAC} and directional statistics,'' \emph{IEEE/ACM Transactions on Audio,
  Speech and Language Processing}, vol.~22, no.~12, pp. 2233--2243, 2014.

\bibitem{winter2007map}
S.~Winter, W.~Kellermann, H.~Sawada, and S.~Makino, ``Map-based underdetermined
  blind source separation of convolutive mixtures by hierarchical clustering
  and l 1-norm minimization,'' \emph{EURASIP Journal on Applied Signal
  Processing}, vol. 2007, no.~1, pp. 81--81, 2007.

\bibitem{sawada2011underdetermined}
H.~Sawada, S.~Araki, and S.~Makino, ``Underdetermined convolutive blind source
  separation via frequency bin-wise clustering and permutation alignment,''
  \emph{IEEE Transactions on Audio, Speech, and Language Processing}, vol.~19,
  no.~3, pp. 516--527, 2011.

\bibitem{abrard2005time}
F.~Abrard and Y.~Deville, ``A time--frequency blind signal separation method
  applicable to underdetermined mixtures of dependent sources,'' \emph{Signal
  Processing}, vol.~85, no.~7, pp. 1389--1403, 2005.

\bibitem{bioucas2013hyperspectral}
J.~M. Bioucas-Dias, A.~Plaza, G.~Camps-Valls, P.~Scheunders, N.~Nasrabadi, and
  J.~Chanussot, ``Hyperspectral remote sensing data analysis and future
  challenges,'' \emph{IEEE Geoscience and Remote Sensing Magazine}, vol.~1,
  no.~2, pp. 6--36, 2013.

\bibitem{ma2014signal}
W.-K. Ma, J.~M. Bioucas-Dias, T.-H. Chan, N.~Gillis, P.~Gader, A.~J. Plaza,
  A.~Ambikapathi, and C.-Y. Chi, ``A signal processing perspective on
  hyperspectral unmixing: Insights from remote sensing,'' \emph{IEEE Signal
  Processing Magazine}, vol.~31, no.~1, pp. 67--81, 2014.

\bibitem{fu2015blind}
X.~Fu, W.-K. Ma, K.~Huang, and N.~D. Sidiropoulos, ``Blind separation of
  quasi-stationary sources: Exploiting convex geometry in covariance domain.''
  \emph{IEEE Transactions Signal Processing}, vol.~63, no.~9, pp. 2306--2320,
  2015.

\bibitem{gannot2001signal}
S.~Gannot, D.~Burshtein, and E.~Weinstein, ``Signal enhancement using
  beamforming and nonstationarity with applications to speech,'' \emph{IEEE
  Transactions on Signal Processing}, vol.~49, no.~8, pp. 1614 --1626, Aug.
  2001.

\bibitem{cohen2004relative}
I.~Cohen, ``Relative transfer function identification using speech signals,''
  \emph{IEEE Transactions on Speech and Audio Processing}, vol.~12, no.~5, pp.
  451--459, 2004.

\bibitem{madhu2008scalable}
N.~Madhu and R.~Martin, ``A scalable framework for multiple speaker
  localization and tracking,'' in \emph{International Workshop for Acoustic
  Echo Cancellation and Noise Control (IWAENC)}, 2008.

\bibitem{dorfan2015tree}
Y.~Dorfan and S.~Gannot, ``Tree-based recursive expectation-maximization
  algorithm for localization of acoustic sources,'' \emph{IEEE Transactions on
  Audio, Speech and Language Processing}, vol.~23, no.~10, pp. 1692--1703,
  2015.

\bibitem{souden2013multichannel}
M.~Souden, S.~Araki, K.~Kinoshita, T.~Nakatani, and H.~Sawada, ``A multichannel
  {MMSE}-based framework for speech source separation and noise reduction,''
  \emph{IEEE Transactions on Audio, Speech, and Language Processing}, vol.~21,
  no.~9, pp. 1913--1928, 2013.

\bibitem{ito2015permutation}
N.~Ito, S.~Araki, and T.~Nakatani, ``Permutation-free clustering of relative
  transfer function features for blind source separation,'' in \emph{23st
  European Signal Processing Conference (EUSIPCO)}, Nice, France, Sep. 2015,
  pp. 409--413.

\bibitem{boardman1993automating}
J.~W. Boardman, ``Automating spectral unmixing of aviris data using convex
  geometry concepts,'' in \emph{The 4th Annual JPL Airborne Geoscience
  Workshop}, 1993, pp. 11--14.

\bibitem{winter1999n}
M.~E. Winter, ``N-findr: An algorithm for fast autonomous spectral end-member
  determination in hyperspectral data,'' in \emph{SPIE's International
  Symposium on Optical Science, Engineering, and Instrumentation}.\hskip 1em
  plus 0.5em minus 0.4em\relax International Society for Optics and Photonics,
  1999, pp. 266--275.

\bibitem{nascimento2005vertex}
J.~M. Nascimento and J.~M. Dias, ``Vertex component analysis: A fast algorithm
  to unmix hyperspectral data,'' \emph{IEEE Transactions on Geoscience and
  Remote Sensing}, vol.~43, no.~4, pp. 898--910, 2005.

\bibitem{araujo2001successive}
M.~C.~U. Ara{\'u}jo, T.~C.~B. Saldanha, R.~K.~H. Galvao, T.~Yoneyama, H.~C.
  Chame, and V.~Visani, ``The successive projections algorithm for variable
  selection in spectroscopic multicomponent analysis,'' \emph{Chemometrics and
  Intelligent Laboratory Systems}, vol.~57, no.~2, pp. 65--73, 2001.

\bibitem{hadad2014multichannel}
E.~Hadad, F.~Heese, P.~Vary, and S.~Gannot, ``Multichannel audio database in
  various acoustic environments,'' in \emph{International Workshop on Acoustic
  Signal Enhancement (IWAENC)}, 2014, pp. 313--317.

\bibitem{vincent2006performance}
E.~Vincent, R.~Gribonval, and C.~F{\'e}votte, ``Performance measurement in
  blind audio source separation,'' \emph{IEEE Transactions on Audio, Speech,
  and Language Processing}, vol.~14, no.~4, pp. 1462--1469, 2006.

\bibitem{stewart1990matrix}
G.~W. Stewart and J.-g. Sun, \emph{Matrix perturbation theory}, 1st~ed.\hskip
  1em plus 0.5em minus 0.4em\relax Academic Press, Jul. 1990.

\end{thebibliography}

\end{document}